\documentclass[12pt,preprint]{emulateapj}
\usepackage{epsfig}
\newcommand{\pmpc}{\hbox {\rm Mpc}^{-1}}

\def\msun{\,{\rm M_\odot}}
\def\kms{\,{\rm km\, s^{-1}}}

\def\himpc{\,h^{-1}{\rm Mpc}}

\shorttitle{Dark Halo Growth and Structure}
\shortauthors{Zhao et al.}
\begin{document}
%
\title{Accurate universal models for the mass accretion histories
and concentrations of dark matter halos}

\author{D.H. Zhao${^{1,2}}$, Y.P. Jing${^{1,2}}$,
H.J. Mo${^{3}}$,  G. B\"orner${^{2}}$} \altaffiltext{1}{Key
Laboratory for Research in Galaxies and Cosmology, Shanghai
Astronomical Observatory, CAS, 80 Nandan Road, Shanghai 200030,
China; e-mail: dhzhao@shao.ac.cn}
\altaffiltext{2}{Max-Planck-Institut f\"ur Astrophysik,
Karl-Schwarzschild-Strasse 1, 85748 Garching, Germany}
\altaffiltext{3}{Department of Astronomy, University of
Massachusetts, Amherst, MA 01003, USA}

%
%
\begin{abstract}
 A large amount of observations have constrained cosmological
parameters and the initial density fluctuation spectrum to a very
high accuracy. However, cosmological parameters change with time and
the power index of the power spectrum varies with mass scale
dramatically in the so-called concordance $\Lambda$CDM cosmology.
Thus, any successful model for its structural evolution should work
well simultaneously for various cosmological models and different
power spectra.

 We use a large set of high-resolution $N$-body simulations of a variety of structure
formation models (scale-free, standard CDM, open CDM, and
$\Lambda$CDM) to study the mass accretion histories, the mass and
redshift dependence of concentrations and the concentration
evolution histories of dark matter halos. We find that there is
significant disagreement between the much-used empirical models in
the literature and our simulations. Based on our simulation results,
we find that the mass accretion rate of a halo is tightly correlated
with a simple function of its mass, the redshift, parameters of the
cosmology and of the initial density fluctuation spectrum, which
correctly disentangles the effects of all these factors and halo
environments. We also find that the concentration of a halo is
strongly correlated with the universe age when its progenitor on the
mass accretion history first reaches 4\% of its current mass.
According to these correlations, we develop new empirical models for
both the mass accretion histories and the concentration evolution
histories of dark matter halos, and the latter can also be used to
predict the mass and redshift dependence of halo concentrations.
These models are accurate and universal: the same set of model
parameters works well for different cosmological models and for
halos of different masses at different redshifts, and in the
$\Lambda$CDM case the model predictions match the simulation results
very well even though halo mass is traced to about 0.0005 times the
final mass, when cosmological parameters and the power index of the
initial density fluctuation spectrum have changed dramatically. Our
model predictions also match the PINOCCHIO mass accretion histories
very well, which are much independent of our numerical simulations
and of our definitions of halo merger trees. These models are also
simple and easy to implement, making them very useful in modeling
the growth and structure of dark matter halos. We provide appendices
describing the step-by-step implementation of our models. A
calculator which allows one to interactively generate data for any
given cosmological model is provided at {\tt
http://www.shao.ac.cn/dhzhao/mandc.html}, together with a
user-friendly code to make the relevant calculations and some tables
listing the expected concentration as a function of halo mass and
redshift in several popular cosmological models. We explain why
$\Lambda$CDM and open CDM halos on nearly all mass scales show two
distinct phases in their mass growth histories. We discuss
implications of the universal relations we find in connection to the
formation of dark matter halos in the cosmic density field.
\end{abstract}

\keywords{cosmology: miscellaneous --- galaxies: clusters: general
--- methods: numerical}

\section{Introduction}

  In the current cold dark matter (CDM) paradigm of structure formation,
a key concept in the buildup of structure in the universe is the
formation of dark matter halos. These halos are quasi-equilibrium
systems of CDM particles formed through nonlinear gravitational
collapse in the cosmic density field. Since galaxies and other
luminous objects are believed to form by cooling and condensation of
the baryonic gas in potential wells of dark matter halos, the
understanding of the formation and properties of CDM halos is an
important part of galaxy formation.

 One of the most important properties of the halo population is their
density profiles. Based on $N$-body simulations, Navarro, Frenk, \&
White (1997, hereafter NFW) found that CDM halos can in general be
approximated by a two-parameter profile,
\begin{equation}
\label{eq:nfw} \rho(r) =
\frac{4\rho_s}{(r/R_s)\left(1+r/R_s\right)^2}\,,
\end{equation}
where $R_s$ is a characteristic ``inner'' radius at which the
logarithmic density slope is $-2$, and $\rho_s$ is the density at
$R_s$. A halo is often defined so that the mean density $\rho_h$
within the halo radius $R_h$ is a factor $\Delta_h$  times the mean
density of the universe (${\bar \rho}$) at the redshift ($z$) in
consideration. The halo mass can then be written as
\begin{equation}
\label{eq:mh}
M_h \equiv \frac{4 \pi}{3} \Delta_h {\bar \rho} R_h^3\,.
\end{equation}
One can define the circular velocity of a halo as
$V_h=(GM_h/R_h)^{1/2}$, so that
\begin{equation}\label{eq:vh}
M_h={V_h^2 R_h\over G}
   ={2^{1/2} V_h^3\over [\Delta_h\Omega_m(z)]^{1/2} H(z)}\,,
\end{equation}
where $H(z)$ is the Hubble parameter and $\Omega_m(z)$ is the cosmic
mass density parameter at redshift $z$. The shape of an NFW profile
is usually characterized by a concentration parameter $c$, defined
as $c \equiv R_h/R_s$. It is then easy to show that
\begin{equation}
\label{eq:rhosc}
  \rho_s=\rho_h \frac {c^3}{12\left[\ln (1+c)-c/(1+c)\right]}\,.
\end{equation}
We denote the mass within $R_s$ by $M_s$, and the circular velocity
at $R_s$ by $V_s$. These quantities are related to $c$ and
$M_h$ by
\begin{equation}
\label{eq:msc}
M_s={\ln 2-1/2\over \ln(1+c)-c/(1+c)}M_h\,,
~~~~~
V_s^2=V_h^2 {cM_s\over M_h}\,.
\end{equation}

 In the literature, a number of definitions have been used for
$\Delta_h$ (hence $R_h$). Some authors opt to use $\Delta_h=200$
(e.g., Jenkins et al. 2001) or $\Delta_h=200/\Omega_m(z)$ (e.g.,
NFW), while others (e.g., Bullock et al. 2001; Jing \& Suto 2002)
choose $\Delta_h=\Delta_{\rm vir}$ according to the spherical
virialization criterion (Kitayama \& Suto 1996; Bryan \& Norman
1998). These different definitions can lead to sizable differences
in $c$ for a given halo, and the differences are
cosmology-dependent. In our discussion in the main text we use
$\Delta_h=\Delta_{\rm vir}$, and adopt the fitting formulae of
$\Delta_{\rm vir}$ obtained by Bryan \& Norman (1998). The value of
$\Delta_{\rm vir}$ ranges from $\sim 180$ at high redshift to $\sim
340$ at present for the current `concordance' $\Lambda$CDM
cosmology. In appendices A and B, we also provide results based on
the other two definitions of $\Delta_h$. To avoid confusing, we
denote halo mass, radius, concentration and circular velocity by
$M$, $R$, $c$, and $V$ for the last definition
($\Delta_h=\Delta_{\rm vir}$), by $M_{\rm 200c}$, $R_{\rm 200c}$,
$c_{\rm 200c}$, and $V_{\rm 200c}$ for the second definition
($\Delta_h=200/\Omega_m(z)$), and by $M_{\rm 200m}$, $R_{\rm 200m}$,
$c_{\rm 200m}$, and $V_{\rm 200m}$ for the first definition
($\Delta_h=200$).

The structure of a halo is expected to depend not only on cosmology
and fluctuation power spectrum, but also on its formation history.
Attempts have therefore been made to relate the halo concentration
to quantities that characterize the formation of the halo. In their
original paper, NFW suggested that the characteristic density of a
halo, $\rho_s$, is a constant ($k$) times the mean density of the
universe, ${\bar \rho (z_f)}$, at the redshift $z_f$ (referred to as
the formation time of the halo by NFW) where half of the halo's mass
was first assembled into progenitors more massive than $f$ times the
halo mass.  NFW used the extended Press--Schechter formula to
calculate $z_f$ and found that the anticorrelation between $c$ and
$M$ observed in their simulation can be reproduced reasonably well
with a proper choice of values of the parameters $k$ and $f$.

  Subsequent investigations demonstrated that additional complexities
are involved in the CDM halo structure. First, halos of a fixed mass
may have significant scatter in their $c$ values (e.g., Jing 2000),
although there is a mean trend of $c$ with $M$. If this trend is
indeed due to a correlation between concentration and formation
time, the scatter in $c$ may reflect the expected scatter in the
formation history for halos of a given mass (e.g., Jing 2000, Lu et
al. 2006). Second, Bullock et al. (2001, hereafter B01) and Eke et
al. (2001, hereafter E01) found that the halo concentration at a
fixed mass is systematically lower at higher redshift. B01 proposed
a model with $c \propto (M/M_\star)^{-0.13}/(1+z)$ ($M_\star$ being
the mass at which the rms of the linear density field is $\sim 1$),
which has a stronger halo-mass dependence, and a much stronger
redshift dependence, than that predicted by the NFW model. E01
proposed a similar model that has a weaker mass dependence, and a
slightly weaker redshift dependence, than the B01 model. Both of
these models have been widely used in the literature to predict the
concentration--halo mass relation. Using the same set of numerical
simulations as that used in B01, Wechsler et al. (2002; hereafter
W02) found that, over a large mass range, the mass accretion
histories (hereafter MAHs) of individual halos identified at
redshift $z_{\rm obs}$ can be approximated by a one-parameter
exponential form,
\begin{equation}
\label{eq:w02mh}
 M(z)=M(z_{\rm obs}){\rm exp}[-2(z-z_{\rm obs})/(1+z_f)]\,,
\end{equation}
where $z_f$ is the formation time of a halo, determined by fitting
the simulated MAH with the above formula. Assuming that $c$ equals
$4.1$ at $z_f$ and grows proportionally to the scale factor
$a\propto 1/(1+z)$, W02 proposed a recipe to predict $c$ for
individual halos through their MAHs, and found that their model can
reproduce the dependence of $c$ on both halo mass and redshift found
in B01. Using a large set of Monte Carlo realizations of the
extended Press--Schechter formalism (EPS), van den Bosch (2002)
showed that the average MAH of CDM halos follows a simple universal
function with two scaling variables. Based on a set of
high-resolution $N$-body simulations, Zhao et al. (2003a, 2003b,
hereafter Z03a and Z03b, respectively) found that the MAH of a halo
in general consists of two distinct phases: an early fast phase and
a late slow phase (see also Li et al. 2007 and Hoffman et al. 2007).
As shown in Z03a, the early phase is characterized by rapid halo
growth dominated by major mergers, which effectively reconfigure and
deepen the gravitational potential wells and cause the collisionless
dark matter particles to undergo dynamical relaxation and mix up
sufficiently to form the core structure, while the late phase is
characterized by slower quiescent growth predominantly through
accretion of material onto the halo outskirt, little affecting the
inner structure and potential. Z03a proposed that the concentration
evolution of a halo depends much on its mass accretion rate and the
faster the mass grows, the slower the concentration increases. In
particular, they predicted that halos which are still in the fast
growth regime should all have a similar concentration, $c\sim 4$.
Using a combination of $N$-body simulations of different
resolutions, Z03b studied in detail how the concentrations of CDM
halos depend on halo mass at different redshifts. They confirmed
that halo concentration at the present time depends strongly on halo
mass, but their results also show marked differences from the
predictions of the B01 and E01 models. The mass dependence of halo
concentrations becomes weaker at higher redshifts, and at $z>3$
halos with $M > 10^{11}h^{-1}{\rm M}_\odot$ all have a similar
median concentration, $c\sim 4$. While the median concentrations of
low-mass halos grow significantly with time, those of massive halos
change very little with redshift. These results are in good
agreement with the empirical model proposed by Z03a and favored by
the {\it Chandra} observation of \citet{SA07}, but are very
different from the predictions of the models proposed by B01, E01
and W02. Recently, Gao et al. (2008) and Macci\'o et al. (2008)
confirmed the results of Z03b with the use of a large set of
simulated halos, suggesting again that the much-used B01 and E01
models need to be revised.

 There have been attempts to modify the early empirical models of halo MAH
and halo concentration to accommodate the new simulation results.
Miller et al. (2006) and Neistein et al. (2006) provided MAH models
based on theoretical EPS formalism. With help of the Millennium
simulation, Neistein \& Dekel (2008) presented a empirical algorithm
for constructing halo merger trees, which is significantly better
than their previous method based on EPS but fails to reproduce the
non-Markov features of merger trees. Moreover, because of the
limited resolution of the simulation, this algorithm is tuned to
reproduce MAHs within redshift 2.5 for only massive halos. McBride
et al. (2009) examined halo MAHs of the Millennium simulation and
modeled the scatter among different halos by fitting individual MAHs
with a two-parameter function, which was proved to be more accurate
than the one-parameter function of W02. Note that both the two
models above are restricted to the specific cosmology and the
specific fluctuation power spectrum of the Millennium simulation,
similar to WMAP1 results. Gao et al. (2008) confirmed the finding of
B01 that the NFW prescription overpredicts the halo concentrations
at high redshift, and tried to overcome this shortcoming by
modifying the definition of halo formation time. However, the
revised model still fails to reproduce the evolution of
concentrations of galaxy sized halos. Modifying the parameters of
the B01 and E01 models can also reduce the discrepancy between these
models with simulation results at high masses, but the overly rapid
redshift evolution of the concentration remains. Motivated partly by
Z03a and Z03b, Macci\'o et al. (2008) presented a model based on
some modification of the B01 model, and found that it can reproduce
the concentration--mass relation at $z=0$ in their simulations.
Unfortunately, their model is not universal, because the
normalization of the concentration--mass relation has to be
calibrated for each cosmological model and for each redshift and
thus the model {\it cannot} be used to predict correctly the
redshift evolution of halo concentrations. In addition, as we will
show below, the model of Macci\'o et al. has the same shortcoming as
the original B01 model in that it predicts too steep a
concentration--mass relation for halos at high redshift in the
current $\Lambda$CDM models and for halos at all redshift in the
`standard' CDM model with $\Omega_{m,0}=1$. Thus, all the existing
empirical models for the halo concentration can at best provide
reliable predictions for dark matter halos in limited ranges of halo
mass and redshift (often around $z=0$), and only for some specific
cosmological models (according to which the empirical models are
calibrated). Clearly, they are insufficient in the era of precision
cosmology.

 We believe that even though a large amount of observations have
constrained cosmological parameters and the initial density
fluctuation spectrum to a very high accuracy, cosmological
parameters change with time and the power index of the power
spectrum varies with mass scale dramatically in the so-called
concordance $\Lambda$CDM cosmology and thus any successful model for
its structural evolution should work well simultaneously for various
cosmological models and different power spectra.

 In this paper, we use $N$-body simulations of a variety of
structure formation models,  including a set of scale-free (SF)
models, two $\Lambda$CDM (LCDM) models, the standard CDM (SCDM)
model, and an open CDM (OCDM) model, to investigate in detail the
MAHs, the mass and redshift dependence of concentrations and the
concentration evolution histories of dark matter halos. We show that
early empirical models for halo MAHs and concentrations all fail
significantly to describe the simulation results. Based on our
simulation results, we develop new empirical models for both the
MAHs and the concentration evolution histories of dark matter halos,
and the latter can also be used to predict the mass and redshift
dependence of halo concentrations. These models are accurate and
universal, in the sense that the same set of model parameters works
well for different cosmological models and for halos of different
masses and at different redshifts. These models are also simple and
easy to implement, making them very useful in modeling the formation
and structure of dark matter halos. Furthermore, the universal
relations we find may also provide important insight into the
formation processes of dark matter halos in the cosmic density
field.

 The organization of the paper is as follows. In Section
\ref{sec_simulations}, we present the set of $N$-body simulations
used in the paper. We describe the simulated MAHs and the
corresponding modeling in Section \ref{sec_MAHs}. Our modeling of
halo concentrations is presented in Section
\ref{sec_concentrations}. Finally, we discuss and summarize our
results in Section \ref{sec_conclusions}. We provide appendices
describing the step-by-step implementation of our models. A
calculator which allows one to interactively generate data for any
given cosmological model is provided at {\tt
http://www.shao.ac.cn/dhzhao/mandc.html}, together with a
user-friendly code to make the relevant calculations and some tables
listing the expected concentration as a function of halo mass and
redshift in several currently popular models of structure formation.

\section{Numerical simulations and dark matter halos}
\label{sec_simulations}

\begin{figure}
\epsscale{1.0} \plotone{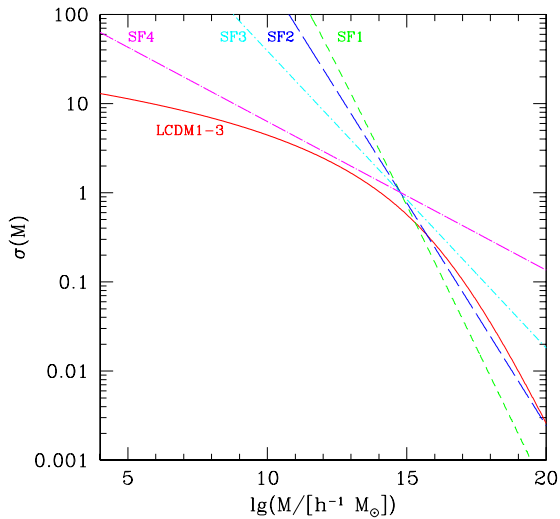} \caption{rms of the linear density
fluctuation $\sigma (M)$ as a function of the mass scale $M$ for
scale-free simulations with $n=1,\,0,\,-1$, and $-2$, and for
$\Lambda$CDM simulations LCDM1--3. \label{fig:sigmamass}}
\end{figure}

We use a very large set of cosmological simulations to study the
formation and structure of dark matter halos of different masses in
different cosmological models. The first subset contains the
simulations that were used in Z03b: the then `concordance'
$\Lambda$CDM model with density parameter $\Omega_{m,0}=0.3$ and a
cosmological constant given by $\Omega_{\Lambda,0}=0.7$. These
simulations, labeled as LCDM1--3 in Table 1, were generated with a
parallel-vectorized Particle-Particle/Particle-Mesh code (Jing \&
Suto 2002). The linear power spectrum has a shape parameter
$\Gamma=\Omega_{m,0} h=0.20$, and an amplitude specified by
$\sigma_8=0.9$, where $h$ is the Hubble constant in units of $100
\kms \pmpc$, and $\sigma_8$ is the rms of the linear density field
smoothed  within a sphere of radius $8\himpc$ at the present time.
We used $256^3$ particles for the simulation of boxsize
$L=25\himpc$, and $512^3$ particles for the other two simulations,
$L=100$ and $300\himpc$ (see Table 1). The simulations with
$L=25\himpc$ and $100\himpc$ were evolved by 5000 time steps with a
force softening length $\eta$ (the diameter of the S2 shaped
particles, Hockney \& Eastwood 1981) equal to $2.5 h^{-1}{\rm kpc}$
and $10 h^{-1}{\rm kpc}$, respectively. In order to examine the
dependence on cosmological parameters, we also use simulations for
the `Standard' CDM model ($\Omega_{m,0}=1$) and the open CDM model
($\Omega_{m,0}=0.3$, $\Omega_{\Lambda,0}=0$), as described in
\citet{Jingsuto02,Jingsuto98}. These are listed as SCDM1, SCDM2 and
OCDM1 in Table 1, together with the corresponding model and
simulation parameters. In addition, we also use a set of four
scale-free (SF) simulations in an Einstein de Sitter universe, with
the linear power-law power spectra given by $P(k)\propto k^n$. These
models are listed as SF1--4 in Table 1 and Figure
\ref{fig:sigmamass} presents the rms of the linear density field
$\sigma (M)$ as a function of the mass scale $M$ for SF1--4 and
LCDM1--3.

We use both the friends of friends (hereafter FOF) algorithm and the
spherical overdensity (hereafter SO) algorithm of Lacey \& Cole
(1994) to identify dark matter groups in the simulations. The FOF
algorithm connects every pair of particles closer than 0.2 times the
mean particle distance and the SO algorithm selects spherical
regions whose average density is equal to $\Delta_{\rm vir}$ times
the mean cosmic density ${\bar \rho}$. We select groups at a total
of 20--30 snapshots for each CDM simulation and at 10 snapshots for
each SF simulation. The outputs are uniformly spaced in the
logarithm of the cosmic scale factor $a$. For each group, we choose
the particle of the highest local density as the group center, and
around it we select as `halo' a spherical region whose average
density is $\Delta_{\rm vir} {\bar \rho}$, with a routine very
similar to the SO algorithm except that here the group center is
fixed. We find that the results presented in this paper do not
depend significantly on the group finding algorithm. Our following
presentation is based on the FOF groups and we use all FOF groups
without applying any further selection criteria.

\begin{table*}
\caption{A Summary of Simulation Parameters\label{tab:sim}}
\begin{center}
\begin{tabular}{ccccccccccc}
\tableline \tableline Simulation& $n_i$ & $h$ & $\sigma_8$ &
$\Omega_{m,0}$ &
$\Omega_{\Lambda,0}$ & $N_p$ & $L[h^{-1}\rm Mpc]$  & $N_{NQ}$ & $\eta$ & $z_i$ \\
\tableline
SF1 & 1 &       &   1   &   1   &   0   &   $512^3$  &  160 & 512 & 0.016  & 1481.3 \\
SF2 & 0 &       &   1   &   1   &   0   &   $512^3$  &  160 & 512 & 0.016  & 363.2  \\
SF3 & -1 &      &   1   &   1   &   0   &   $512^3$  &  160 & 512 & 0.016  & 75.0   \\
SF4 & -2 &      &   1   &   1   &   0   &   $512^3$  &  160 & 512 & 0.016  & 11.3   \\
LCDM1   & 1 &   0.667  &  0.9  &  0.3 &   0.7  &  $256^3$  &  25  & 512 & 0.0025 & 72    \\
LCDM2   & 1 &   0.667  &  0.9  &  0.3 &   0.7  &  $512^3$  &  100 & 512 & 0.01   & 72    \\
LCDM3   & 1 &   0.667  &  0.9  &  0.3 &   0.7  &  $512^3$  &  300 & 512 & 0.03   & 36    \\
LCDM4   & 1 &   0.71 &  0.85  &  0.268 &   0.732  &  $1024^3$  &  300  & 1024 & 0.01 & 72    \\
LCDM5   & 1 &   0.71 &  0.85  &  0.268 &   0.732  &  $1024^3$  &  1200 & 1024 & 0.072   & 72    \\
LCDM6   & 1 &   0.71 &  0.85  &  0.268 &   0.732  &  $1024^3$  &  1800 & 1024 & 0.144   & 72    \\
OCDM1   & 1 &   0.667  &  1.0  &  0.3 &   0   &   $256^3$  &  50  & 256 & 0.02   & 72    \\
SCDM1   & 1 &   0.5  &  0.55 &  1  &    0   &   $256^3$  &  25  & 512 & 0.0025 & 72    \\
SCDM2   & 1 &   0.5  &  0.55 &  1  &    0   &   $512^3$  &  100 & 256 & 0.02   & 72    \\
SCDM3   & 1 &   0.5  &  0.55 &  1  &    0   &   $512^3$  &  300 & 256 & 0.03   & 36    \\
\tableline
\end{tabular}
\tablecomments{The Bardeen et al. (1986) fitting formula for
transfer function is used for all CDM simulations except LCDM4--6.
For LCDM4--6, the fitting formula of Eisenstein \& Hu (1998) is
adopted, with baryon density parameter $\Omega_{b,0}=0.045$. Note
that a short-wavelength cutoff in the linear power spectrum must be
included in the SF1 simulation.}
\end{center}
\end{table*}

The simulations described above are also used to study the density
profiles and the concentrations of dark matter halos. We use all
halos containing more than 500 particles for our analysis. Each halo
is fitted with the NFW profile, using a similar method as described
in Z03b. As demonstrated in Z03b, 500 particles are sufficient for
measuring the concentration parameter reliably. In order to examine
concentrations of very massive halos, we also use three additional
simulations with large boxsizes \citep[see][]{Jingsutomo07}. These
simulations are listed as LCDM4--6 in Table 1. Note that the model
parameters of these simulations are slightly different from those of
LCDM1--3. Furthermore, the initial power spectrum for these
simulations is obtained using the fitting formula of \citet{EH98},
which includes the effects of baryonic oscillations. The details of
these simulations can be found in \citet{Jingsutomo07}, and all the
change in model parameters are properly taken into account when we
compare our model predictions of halo concentrations with the
simulation results. For the same reason, we also use a SCDM
simulation of boxsize $L=300\himpc$ (listed as SCDM3 in Table 1) to
investigate the concentration--mass relation at the highest halo
mass end in this model.

\section{A universal model for the halo mass accretion histories}
\label{sec_MAHs}

Since we want to study how a halo grows with time, we need to
construct the main branch of the merger tree for each halo.  Given a
group of dark matter particles at a given output time (which we
refer to as group 2), we trace all its particles back to an earlier
output time. A group (group 1) at the earlier output is selected as
the ``main progenitor'' of group 2 if it contributes the largest
number of particles to group 2 among all groups at this earlier
output.  We found that in most cases more than half of the particles
of group 1 is contained in group 2. We refer to group 2 as the
``main offspring'' of group 1. We construct the main branch of the
merger tree, for each of the most massive 10000 halos identified at
redshift $z=0$, until the number of particles in the progenitor
drops below 10 or the main progenitor cannot be found anymore. More
than $97\%$ of the histories analyzed here can be traced until the
particle number goes below 10.

\begin{figure}
\epsscale{0.6} \plotone{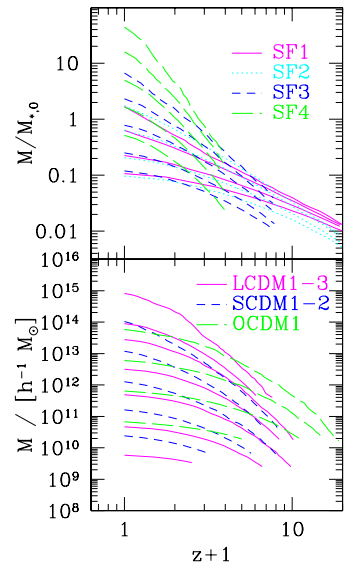} \caption{Median halo MAHs obtained
from scale-free simulations with $n=1,\,0,\,-1,$ and $-2$ and from
CDM simulations LCDM1--3, SCDM1--2, and OCDM1. Different lines
represent median histories with different final halo masses. For
each final mass, the result is plotted to a redshift when the
histories of $90\%$ of the halos can still be traced.
\label{fig:mahsSFsimu}}
\end{figure}

 As an illustration, Figure \ref{fig:mahsSFsimu} shows the median MAHs
of halos both in the SF simulations and in the CDM models. The halo
mass is normalized by the characteristic mass $M_{*,0}$ at the final
output in the SF simulations. Clearly, MAHs depend on the power
spectrum. This is because halos grow faster in a model with a
smaller value of $n$. The effect of cosmological parameters can also
be seen clearly in CDM models: the growth of halos is slower in the
LCDM model than in the SCDM model (despite the fact that the LCDM
has more clustering power than the SCDM) and is the slowest in the
OCDM. In our analysis, we have taken into account the incompleteness
effect of the MAHs at the earliest epochs when the progenitors
containing less than 10 particles cannot be followed in the
simulation. This effect leads to an overestimation of MAHs. In order
to correct for it, statistical analysis of the MAHs is carried out
only up to the redshift where the progenitors of $90\%$ of all the
halos in consideration can be traced.

 As one can see from Figure \ref{fig:mahsSFsimu}, the MAHs show a
complex dependence on power spectrum, cosmology and halo mass. The
goal of this section is to find an order in such complexity so as to
obtain a universal model to describe all the MAHs.

\subsection{The Expected Asymptotic Behavior at High Redshift}
\label{sec_asymptotic}

\begin{figure}
\epsscale{1.0} \plotone{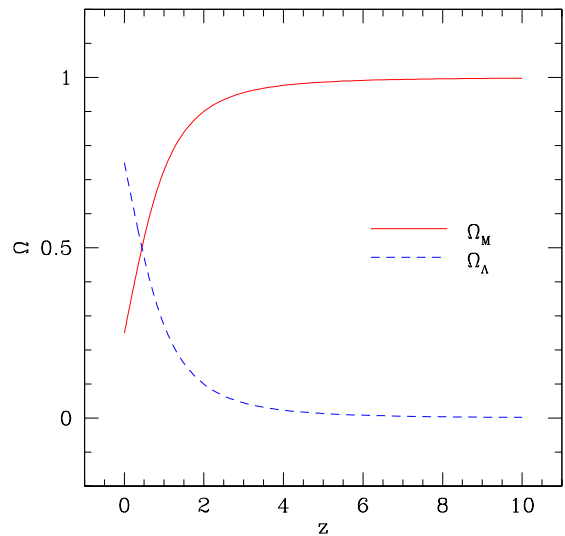} \caption{Evolution of cosmological
density parameter $\Omega_m$ and cosmological constant parameter
$\Omega_\Lambda$ for a flat universe with cosmological constant.
\label{fig:omegaevolution}}
\end{figure}

  Before modeling the MAHs in detail, let us first consider some generic
properties of the growth of CDM halos in the cosmic density field.
In the CDM scenario, structures form in a hierarchical fashion, and
the growth of dark matter halos in general has the following
properties. First, at any given time, more massive halos are, on
average, growing faster because they sit in higher density
environments. Second, the higher the redshift, the faster the halos
of a fixed mass grow because at high redshift they are relatively
more massive with respect to others. Third, the MAH of a halo
depends on power spectrum of the initial density fluctuation, as
mentioned before. Fourth, the MAH of a halo also depends on
cosmology, because of the change of the background expansion. The
last, halo growth also suffers from some nonlinear processes in
their local environments. All these factors entangle together and
keep varying in their own complicated ways. For example, in a
$\Lambda$CDM universe, the cosmological density parameter $\Omega_m$
decrease with time and the power index of the initial density
fluctuation spectrum increase with mass scale dramatically, as shown
in Figure \ref{fig:omegaevolution} and Figure \ref{fig:sigmamass},
respectively. This is why one need a universal evolution model which
works well for various cosmologies and different spectra, as
mentioned in the introduction already. For exactly the same reason,
however, we may not expect that variations of all these factors
compensate exactly to bring us a universal function form for MAH in
terms of $M$ and $z$. Thus, we turn to model halo mass accretion
rate instead, which can be integrated to build the MAH, and in order
to describe the accretion rate accurately and universally, we need
to disentangle all effects due to halo mass, redshift, the
cosmological parameters, the linear power spectrum and those
relevant nonlinear processes.

Consider two average halos of different masses at a given time.
Since the more massive one grows faster, we expect that at a
slightly earlier time, the two halos were closer in mass (in terms
of the mass ratio), and the masses of their progenitors at higher
redshift are expected to be even closer. The two halos are expected
to have similar mass accretion rate at sufficiently high redshift,
because otherwise with finite difference in accretion rates their
mass accretion trajectories would cross. This would lead to an
unlikely situation where {\it more} massive progenitors actually
grow {\it slower} than low-mass progenitors at redshifts higher than
the crossing redshift. This means that the average MAHs of halos of
different masses should all have the same asymptotic behavior at
high redshift. Here we try to look for this behavior.

Let us first consider an Einstein de Sitter universe with an initial
Gaussian density fluctuation field of a power-law power spectrum,
$P(k)\propto k^n$.  As there is no characteristic scale either in
time or in space, this is a case where the structure formation is a
self-similar process. When halo mass is scaled with a time-dependent
characteristic mass, all statistical quantities as functions of the
scaled mass should be the same at all redshifts. Given a linear
power spectrum, we can linearly extrapolate to $z=0$, and estimate,
the rms of the linear density field $\sigma (M)$ on a mass scale
$M$. According to the spherical collapse model, the linear critical
overdensity for collapse at redshift $z$ is $\delta_c(z)=\delta'_c
[\Omega_m(z),\Omega_\Lambda(z)]/D(z)$, where $D(z)$ is the linear
growth factor, which is $1/(1+z)$ in an Einstein de Sitter universe,
and $\delta'_c[\Omega_m(z),\Omega_\Lambda(z)]$ is the conventional
critical overdensity for collapse at redshift $z$, which is a
constant $\cong 1.686$ in this universe. Because $\delta_c(z)$ is a
function of only $z$ and $\sigma(M)$ is a function of only $M$, we
can think of $1/\sigma^2(M)$ as the ``mass'' and $1/\delta^2_c(z)$
as the ``time''. Furthermore, since $M_\star$, which satisfies
$\sigma(M_\star)=\delta_c(z)$, represents the characteristic
nonlinear mass at redshift $z$, $1/\delta^2_c(z)$ actually
corresponds to the characteristic nonlinear ``mass'' at redshift $z$
if $1/\sigma^2(M)$ is regarded as ``mass''. In this case, the scaled
``mass'' is $\nu=\delta^2_c(z)/\sigma^2(M)$, i.e., the peak height
of the halo in the conventional terminology.

  Suppose that there are some halos at a given redshift $z_{\rm obs}$
and the average mass growth rate is
$d\ln\sigma^2(M)/d\ln\delta^2_c(z)|_{z_{\rm obs}}=1$, i.e.,
$d\ln\nu/d \ln\delta^2_c(z)|_{z_{\rm obs}}=0$ so that $\nu={\rm
constant}$. Then at a slightly higher redshift, $z_1$, the scaled
mass $\nu$ of their progenitors will be the same as that at $z_{\rm
obs}$. Since the statistical properties of halos of the same scaled
mass are the same for all redshifts in the self-similar case, these
progenitors will have the same growth rate as that at $z_{\rm obs}$,
i.e., both the mass growth rate
$d\ln\sigma^2(M)/d\ln\delta^2_c(z)|_{z_{\rm obs}}$, and the scaled
mass $\nu$, remains constant with time.  Therefore, the main
progenitors at different redshifts will all have the same $\nu$ and
the same growth rate as they have at redshift $z_{\rm obs}$, and the
average MAH will be a straight line in the logarithmic diagram of
$\sigma$ versus $\delta_c$. This line is exactly the asymptotic
behavior we are looking for.

In the case of cold dark matter cosmology (SCDM, LCDM, and OCDM),
structure formation is not self-similar. Nevertheless, we can still
use $1/\sigma^2(M)$ to represent halo mass, $1/\delta^2_c(z)$ to
represent time, and $\nu$ to represent the scaled mass. However, it
is not guaranteed that the $\sigma$--$\delta_c$ relation has a
unique asymptotic behavior as it has in the SF case. As we will see
below, simulated median MAHs of different final masses show the same
asymptotic behavior at high $z$ not only for the SF models but also
for the CDM models. We can then use the result to guide our modeling
of the halo MAHs in simulations.

\subsection{Toward a Universal Model for the Halo Mass Accretion Histories}

\begin{figure}
\epsscale{1.0} \plotone{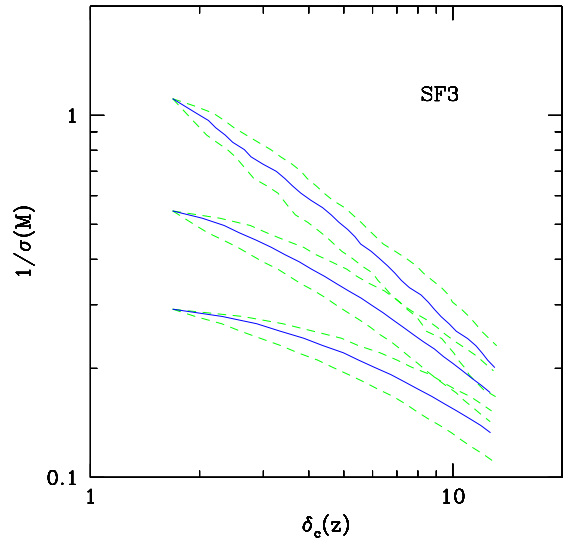} \caption{Median (solid blue) and
scatter (dashed green) of the simulated MAHs, in terms of
$1/\sigma(M)$ vs. $\delta_c(z)$, for halos of three final masses in
the scale-free case with $n=-1$. Dashed green lines enclose $60\%$
of the histories in each mass bin. \label{fig:mahsSF2disp}}
\end{figure}

If we use $1/\sigma(M)$ as mass and $1/\delta_c(z)$ as time, the MAH
looks like the curves shown in Figure \ref{fig:mahsSF2disp}, where
we have adopted the $n=-1$ SF model as an example. The figure
clearly shows the asymptotic behavior described above. The MAHs of
halos of different present masses converge at high redshift, and the
scaled mass $\sqrt{\nu} = \delta_c(z)/\sigma(M)$ approaches a
constant, as one can see in the case of the most massive halos in
the figure, and as expected from the asymptotic behavior discussed
above.

\begin{figure}
\epsscale{1.0} \plotone{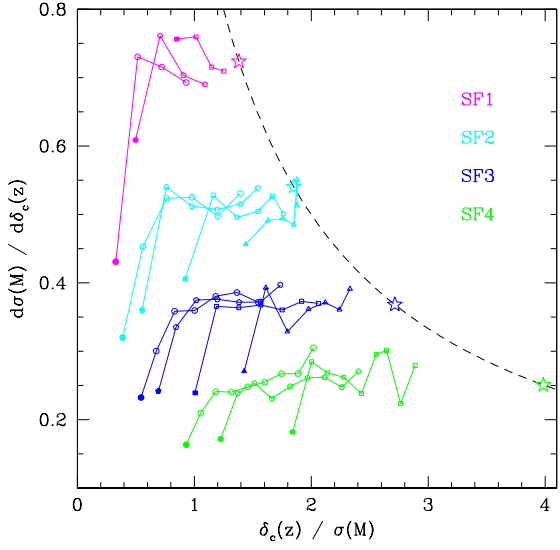} \caption{Median mass accretion rate
for our four scale-free models. Purple, blue, dark blue and green
symbols and lines represent results for scale-free simulations with
$n=1,\,0,\,-1$, and $-2$, respectively. Different symbols represent
median histories with different final halo masses, and in each
history we use a solid symbol to denote the final snapshot of the
history. The four stars and the dashed curve represent the expected
asymptotic behavior for high peaks in scale-free models.
\label{fig:accretionrate1}}
\end{figure}

 Figure \ref{fig:accretionrate1} shows the median
accretion rate, $d\sigma(M)/d\delta_c(z)$, as a function of the
scaled mass, $\sqrt{\nu}=\delta_c(z)/\sigma(M)$, for the four SF
models. For each model, a set of symbols of the same type connected
with a solid line represents accretion rates of progenitors of the
halos of a given mass at $z=0$, and the filled symbol represents the
end of the MAH at $z=0$. It is interesting to note that, for each SF
model, the results for different halo masses are similar except at
the end of the history. The accretion rate plotted in this way is
approximately a constant for all the progenitors. As mentioned
above, in each model the scaled mass is asymptotically a constant at
high $z$. This corresponds to a line, $d\sigma(M)/d\delta_c(z) =
\sigma(M)/\delta_c(z)$, which is plotted as the dashed line in the
figure together with a star showing the corresponding asymptotic
accretion rate.  At the end of each history, the mass accretion rate
is somewhat lower than the average of their progenitors. This
deviation from self-similarity is not a numerical artifact, as the
simulation resolution is sufficient at $z=0$. Instead, it
reflects the fact that the main progenitors are a special subset of
the total halo population. While the halos of a given $\nu$ at the end of MAHs
are chosen only by mass, their progenitors have an additional
selection bias: they are chosen to be main progenitors of more massive
halos. This biases the progenitor halos to have higher mass accretion
rates than typical halos of a given $\nu$ value. In other words,
a fraction of halos in the total
population are {\it not} main progenitors of larger halos in the
future but instead will be swallowed by more massive
halos. Their mass accretion rates may have been suppressed by their
massive neighbors due to environment-heating or tidal-stripping, as
envisaged in Wang et al. (2007), and thus their merger trees should
also have some non-Markov features.

\begin{figure}
\epsscale{1.0} \plotone{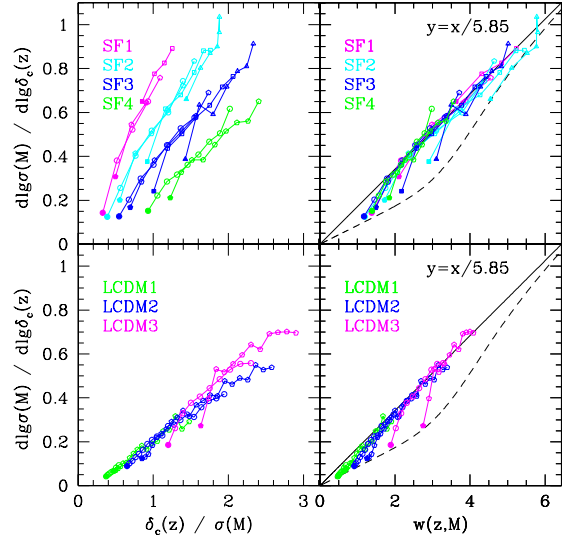} \caption{Same as Figure
\ref{fig:accretionrate1}, except that both the horizontal and
vertical axes have been rescaled as indicated by the axis labels.
Results for the six LCDM histories shown in Figure
\ref{fig:mahsSFsimu} with final halo mass larger than $10^{10}
h^{-1} M_{\odot}$ are also plotted here, in the two lower panels.
The black solid straight lines in the right two panels have a slope
$1/5.85$, and the dashed curve in each of the two panels represents
the envelope given by halos in the end of the MAHs in simulations
LCDM1--3. \label{fig:accretionrate2}}
\end{figure}

  As seen in Figure \ref{fig:accretionrate1}, accretion rates are
different in different SF models, again indicating that the
accretion rate depends on the shape of the power spectrum. In order
to model such dependence, we use $d\ln\sigma(M)/d\ln\delta_c(z)$
instead of $d\sigma(M)/d\delta_c(z)$ to represent the accretion
rate. The results are shown in the upper left panel of Figure
\ref{fig:accretionrate2}, which again indicates that the accretion
rate depends strongly on the power index of the linear spectrum.
However, we find that this dependence can be scaled away if we use
the following variable instead of $\delta(z)/\sigma(M)$:
\begin{equation}\label{eq:wzm}
 w(z,M) \equiv \delta_c(z)/s(M)\,,
\end{equation}
where
\begin{equation}\label{eq:sasM}
s(M) \equiv \sigma(M) \times 10^{d\lg\sigma/{d\lg m}|_M}\,.
\end{equation}
This is shown in the upper right panel of Figure
\ref{fig:accretionrate2}. With the use of $w(z,M)$, all the halo
MAHs lie on top of each other, except for the snapshots close to the
end of the MAHs. Even more remarkably, the accretion rates of the
progenitors at high redshift can all be well described by a straight
line,
\begin{equation}\label{eq:accretionratehigh}
d\lg\sigma(M)/{d\lg\delta_c(z)}=\frac{w(z,M)}{5.85} \,,
\end{equation}
which is shown in the panel as the solid line. Note that $\lg s(M)$
is equal to $\lg\sigma(M)$ plus its logarithmic slope at mass $M$:
$d\lg\sigma/{d\lg m}|_M$. For a given power spectrum with $n>-3$,
$\sigma(M)$ decreases with halo mass and so $w(z,M)$ increases with
halo mass. The simple relation given by Equation
(\ref{eq:accretionratehigh}) also describes well the MAHs at high
redshifts in the LCDM model (see the lower two panels of Figure
\ref{fig:accretionrate2}), although the power spectrum and the
cosmology are very different from the SF models.

\begin{figure}
\epsscale{0.6} \plotone{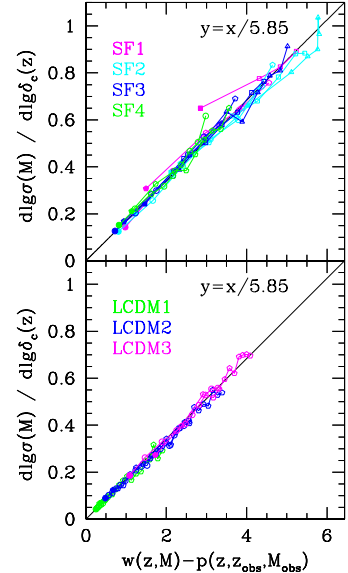} \caption{Same as Figure
\ref{fig:accretionrate2}, except that the horizontal axis is
rescaled again, as indicated by the axis label. Here $p$ is defined
in the text. \label{fig:accretionrate3}}
\end{figure}

 As shown in the two right panels of Figure \ref{fig:accretionrate2},
the accretion rate is systematically lower than that given by
Equation (\ref{eq:accretionratehigh}) at the end of the MAHs. We
find that this decrease of the accretion rate can be accounted for
by replacing $w(z,M)$ with $w(z,M)-p(z, z_{\rm obs}, M_{\rm obs})$.
The shift, $p(z, z_{\rm obs}, M_{\rm obs})$, in the horizontal axis
depends on redshift and halo mass in the following way,
\begin{eqnarray}\label{eq:pz}
p(z, z_{\rm obs}, M_{\rm obs})=p(z_{\rm obs}, z_{\rm obs}, M_{\rm
obs})\nonumber \\
\times {\rm Max} \left[0,1-\frac{\lg\delta_c(z)-\lg\delta_c(z_{\rm
obs})}{0.272/w(z_{\rm obs}, M_{\rm obs})}\right] \,,
\end{eqnarray}
where
\begin{equation}\label{eq:pzobs}
p(z_{\rm obs}, z_{\rm obs}, M_{\rm obs})=\frac{1}{1+[w(z_{\rm obs},
M_{\rm obs})/4]^6}\frac{w(z_{\rm obs}, M_{\rm obs})}{2}\,
\end{equation}
is the shift at $z_{\rm obs}$, the redshift at which the final halo
is identified. The horizontal gap between the dashed and solid
curves in the right two panels of Figure \ref{fig:accretionrate2}
shows this shift. As one can see, the shift describes well the
deviation of the mass accretion rates in the end of the histories
(the solid symbols) for the LCDM model. For the SF models, the shift
appears to be too much, but this is mainly due to the fact that the
interval of the last two snapshots in the SF models is quite big,
and the accretion rate is not estimated accurately. Figure
\ref{fig:accretionrate3} shows $d\lg\sigma(M)/{d\lg\delta_c(z)}$
versus $w(z,M)-p(z, z_{\rm obs}, M_{\rm obs})$. It is clear that the
relation is much tighter, and is well described by the following
relation,
\begin{equation}
\label{eq:accretionrate}
d\lg\sigma(M)/{d\lg\delta_c(z)}=\frac{w(z,M)-p(z, z_{\rm obs},
M_{\rm obs})}{5.85} \,,
\end{equation}
which is shown as the solid lines. The same results are also found
for the SCDM and OCDM simulations, although they are not shown here.

\begin{figure}
\epsscale{1.0} \plotone{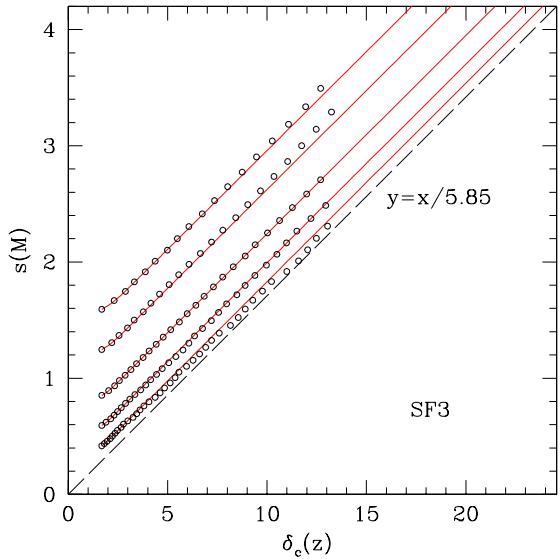} \caption{Predicted and simulated
median halo MAHs for the scale-free model with $n=-1$. Black symbols
are simulation results, and solid red lines represent our model
predictions. Results are shown for halos in five bins of final mass.
\label{fig:mahssigSF3}}
\end{figure}

\begin{figure}
\epsscale{1.0} \plotone{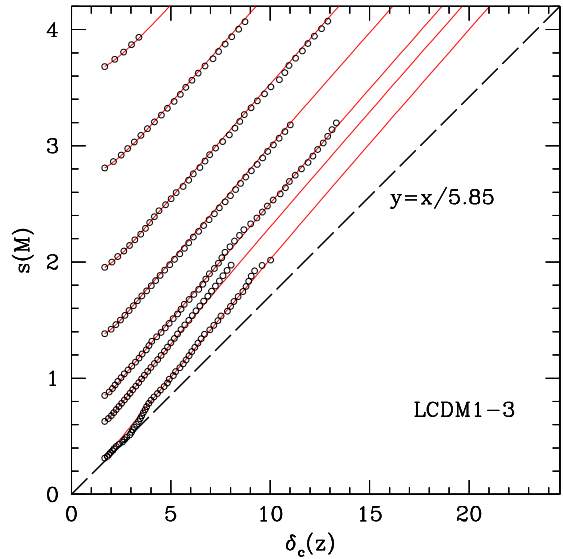} \caption{Predicted and simulated
median halo MAHs for simulations LCDM1--3. Symbols and lines are the
same as in Figure \ref{fig:mahssigSF3}. Here results are shown for
halos in seven bins of final mass. \label{fig:mahssigLCDM123}}
\end{figure}

\begin{figure}
\epsscale{1.0} \plotone{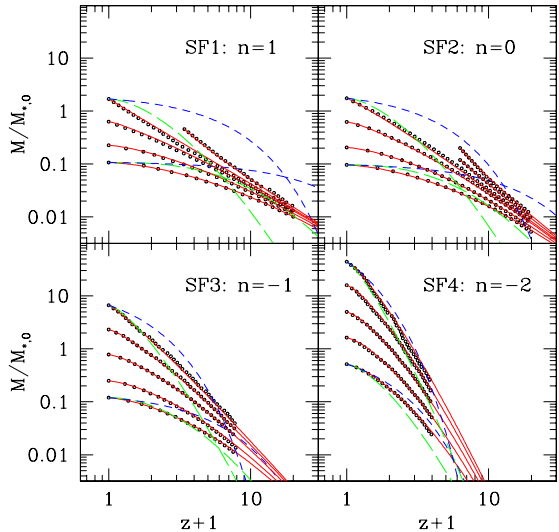} \caption{Predicted and simulated
median halo MAHs for scale-free models with $n=1,\,0,\,-1$ and $-2$.
Black symbols are simulation results, while solid red lines
represent our model predictions. Those histories ended at high
redshifts are MAHs with final halos chosen at those redshifts. The
short-dashed blue lines and the long-dashed green lines are
predictions of the model of W02 and predictions of the model of van
den Bosch (2002; based on Monte Carlo realizations of EPS),
respectively. \label{fig:mahsSF}}
\end{figure}

\begin{figure}
\epsscale{1.0} \plotone{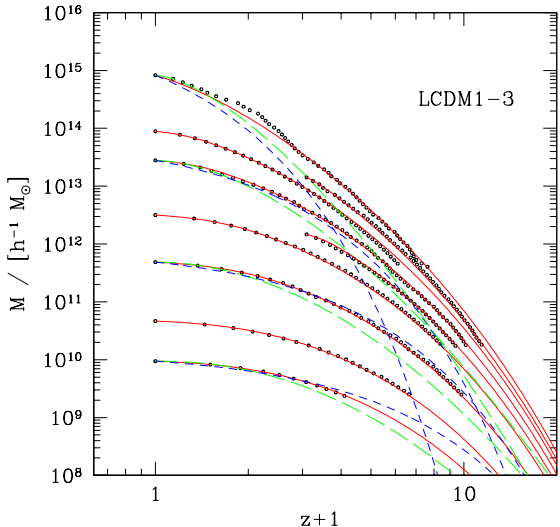} \caption{Predicted and simulated
median halo MAHs for simulations LCDM1--3. Symbols and lines are the
same as in Figure \ref{fig:mahsSF}. \label{fig:mahsLCDM123}}
\end{figure}

\begin{figure}
\epsscale{0.6} \plotone{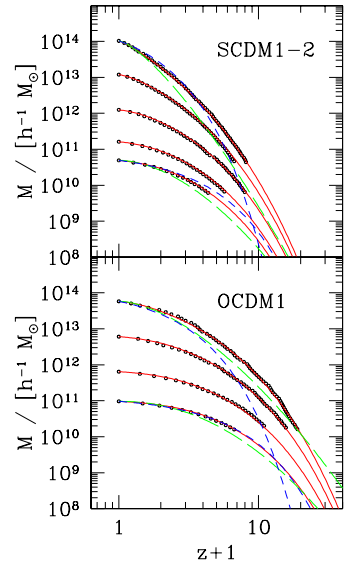} \caption{Predicted and simulated
median halo MAHs for simulations SCDM1--2 and OCDM1. Symbols and
lines are the same as in Figure \ref{fig:mahsSF}.
\label{fig:mahsSCDM12OCDM}}
\end{figure}

 Given a cosmology and a linear power spectrum, it is easy to
calculate $s(M)$ for a halo of mass $M_{\rm obs}$ at a given
redshift $z_{\rm obs}$. From Equation (\ref{eq:accretionrate}), one
can estimate the mass accretion rate,
$d\lg\sigma(M)/{d\lg\delta_c(z)}$ at $z_{\rm obs}$. One can then
compute $\sigma(M)$ [or equivalently $s(M)$] at a redshift $z$
incrementally higher than $z_{\rm obs}$, thus tracing the MAH to
higher redshifts. The MAH, in terms of $s(M)$ and $\delta_c(z)$,
thus computed are shown in Figures \ref{fig:mahssigSF3} and
\ref{fig:mahssigLCDM123}. Since $s(M)$ is a monotonic function of
$M$ and $\delta_c (z)$ a monotonic function of $z$, the
$s(M)$--$\delta_c(z)$ relation can easily be converted into an MAH
in terms of $M$ and $(1+z)$. Figures
\ref{fig:mahsSF}--\ref{fig:mahsSCDM12OCDM} show the median MAHs so
obtained for halos of different final masses at different redshifts
in various models (solid curves), in comparison with the simulation
results (circles). For the $\Lambda$CDM case, the model predictions
match the simulation results very well even though halo mass is
traced to about 0.0005 times the final mass, when cosmological
parameters and the power index of the initial density fluctuation
spectrum have changed dramatically. Moreover, the same set model
parameters works pretty good for all halo masses, all redshifts and
all cosmologies we are studying here. The typical errors of the
predicted median MAHs in most cases are $\leq 10\%$. Somewhat larger
errors seen in the highest mass bin in Figure \ref{fig:mahsLCDM123}
may be due to the inaccurate determination in the simulation because
the number of halos is small in this mass bin. These predictions are
much more accurate than those of the model of W02, shown as the
short-dashed lines in Figures
\ref{fig:mahsSF}--\ref{fig:mahsSCDM12OCDM},\footnote{Following the
instruction on Bullock's Web site
http://www.physics.uci.edu/{\textasciitilde}bullock/CVIR/, we set
$F=0.015$ in the B01 model (see Section \ref{sec_cmzrelation} for
more details) and use the returned collapsing redshift as input for
the free parameter $z_f$ in the W02 model, Equation
(\ref{eq:w02mh}).} and those of the model of \citet{vandenBosch02},
shown as the long-dashed lines in Figures
\ref{fig:mahsSF}--\ref{fig:mahsSCDM12OCDM}.\footnote{Note that
predictions of the model of \citet{vandenBosch02} are the average
MAHs while the predictions of ours are the median MAHs. If the same
definition is adopted, the difference between predictions of these
two models will be even larger in the CDM cases for which our model
always predicts more massive progenitors. This is because individual
MAHs show a log-normal distribution (D. H. Zhao et al. 2010, in
preparation) and so the average value should be somewhat higher than
the median one.} Furthermore, unlike the latter two models, the MAHs
predicted by our model for halos of different final masses do not
cross at high redshift. Note that the W02 model works pretty well
for low-mass halos in the $\Lambda$CDM and the OCDM models
($M<10^{12}M_{\odot}$). However, for more massive halos in these two
models and for all halos in the scale-free and the SCDM simulations,
it fails to provide an accurate description for the MAHs.
Even though the W02 model and ours give similar MAHs for low-mass
halos in some CDM models, they have very different asymptotic
behaviors at high redshift, because the W02 model is an exponential
function of $z$ while ours is a power-law function of $z$.

\begin{figure}
\epsscale{1.0} \plotone{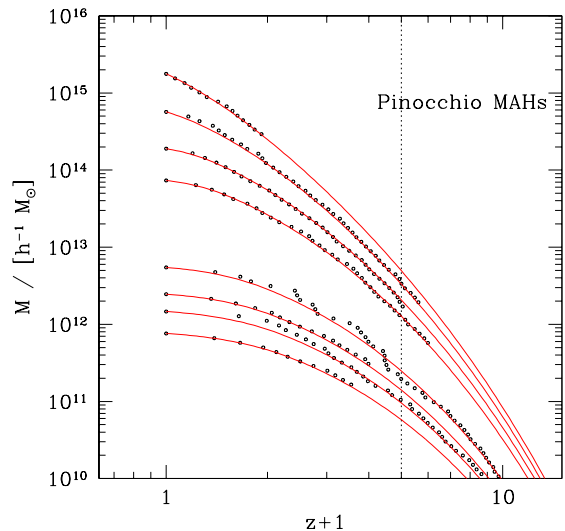} \caption{Predicted and the
PINOCCHIO median halo MAHs for a cosmology model the same as
LCDM1--3. Black symbols represent the median redshifts of the halo
MAHs output by the PINOCCHIO code automatically, while solid red
lines are our model predictions. Four different PINOCCHIO runs with
box sizes of 25, 100, 300, and $1200\himpc$ and particle numbers of
$256^3$, $256^3$, $512^3$, and $256^3$ are analyzed here. The
vertical dotted line indicates the redshift 4.
\label{fig:mahsPINOCCHIO}}
\end{figure}

Several months after this current paper was submitted to the Journal
and the Internet, McBride et al. (2009) modeled the distribution of
halo MAHs of the Millennium simulation by fitting individual MAHs
with a two-parameter function, instead of the one-parameter function
of W02. They found that our model prediction for the mass accretion
rate to have a slightly steeper $z$ dependence than their Equation
(9) where our median value is within $20\%$ of their median mass
accretion rate at $z \sim 0$ but exceeds theirs by a factor of $\sim
2$ at $z \sim 4$. The reason for this large discrepancy is unknown,
however, it is worthy to point out that since we combine a set of
high-resolution simulations, the dynamical ranges explored here are
larger than the Millennium Simulation and our halo samples are well
enough for analyzing the median value. To test our model further, we
utilize the Lagrangian semianalytic code PINOCCHIO,\footnote{A
public version of PINOCCHIO is available from the Web site:
http://adlibitum.oats.inaf.it/monaco/Homepage/Pinocchio.} proposed
by Monaco et al. (2002) for identifying dark matter halos in a given
numerical realization of the initial density field in a hierarchical
universe. Figure \ref{fig:mahsPINOCCHIO} compares our model
predictions with the median redshifts of the MAHs produced by the
PINOCCHIO code, for a cosmology model the same as LCDM1-3 and very
similar to that of the Millennium Simulation. Here four different
PINOCCHIO runs with box sizes of 25, 100, 300, and $1200\himpc$ and
particle numbers of $256^3$, $256^3$, $512^3$, and $256^3$ are
analyzed and, to avoid artifact, only those MAHs which converge
among runs of different boxsizes are plotted. Again, at all
redshifts available, our model predictions match the median values
of the PINOCCHIO MAHs very well in halo mass range probed by the
Millennium Simulation. The fluctuations on the median MAHs in the
smallest box are due to the small number statistics. Note that the
PINOCCHIO MAHs are automatically output by the PINOCCHIO code
without any postprocessing and so are much independent of our
numerical simulations and of our definitions of halo merger trees.

 It is interesting to examine the scaling relations obtained above
more closely. At sufficiently high $z$ when
$\lg\delta_c(z)-\lg\delta_c(z_{\rm obs})>0.272/w(z_{\rm obs})$,
Equation (\ref{eq:pz}) gives $p=0$, and Equation
(\ref{eq:accretionrate}) reduces to a simpler form. In this case,
$w(z, M)\equiv \delta_c(z)/s(M)=5.85$ corresponds to
$d\lg\sigma(M)/{d\lg\delta_c(z)}=1$, or equivalently to
$d\lg\nu/{d\lg\delta^2_c(z)}=0$. The scaled mass, $\sqrt{\nu}$, is
then independent of time, so is $w$ in the SF case. This corresponds
to the high-redshift asymptotic behavior discussed in the last
subsection. For halos more massive than this characteristic scale,
the model predicts that $d\lg\sigma(M)/{d\lg\delta_c(z)}>1$ and that
$\sqrt{\nu}$ increases with time. For scale-free cases, this means
that the most massive halos will grow faster and faster, as
demonstrated in the two upper panels of Figure \ref{fig:mahsSF},
while for $\Lambda$CDM and OCDM cases, this kind of rapidly growing
halos are very rare because neither the $\delta_c$--$1+z$ relation
nor the $\sigma$--$M$ relation is a simple power law. Both the
decrease of cosmological density parameter $\Omega_m$ with time and
the increase of power index of the power spectrum with mass scale
will slow down the halo mass growth rate, as shown above. This is
why at the present time halos on nearly all mass scales show two
distinct phases in their mass growth histories, as found by Z03b and
Z03a.

 For the SF models, we can obtain some useful analytic formulae,
because here $\sigma(M)$ has a simple power-law dependence on $M$:
$\sigma \propto M^{-(n+3)/6}$. First, using the high-$z$ asymptote,
$w(z,M)=5.85$, we have $\delta_c(z)/\sigma(M)=5.85 \times
10^{d\lg\sigma/{d\lg m}}=5.85 \times 10^{-(n+3)/6}$ and $\nu=5.85^2
\times 10^{-(n+3)/3}$. This means that the mass of a median
progenitor at sufficiently high redshift is a fixed fraction of the
characteristic mass: $M=5.85^{6/(n+3)}(M_*/10)$. For $n=1,\,0,\,-1$
and $-2$, these progenitors have a typical mass $M/M_*=1.415, 3.422,
20.02$, and $4008.07$, respectively. Second, because
$d\lg\sigma/d\lg m=-(n+3)/6$ is a constant in an SF model, we have
$d\lg
s(M)/{d\lg\delta_c(z)}=d\lg\sigma(M)/{d\lg\delta_c(z)}=w(z,M)/5.85
=\delta_c(z)/s(M)/5.85$ when $p(z)=0$. The solution is
$s(M)=\delta_c(z)/5.85+C$ with $C$ a constant along each median
history. This solution describes the MAHs in the SF models quite
well, as shown in Figure \ref{fig:mahssigSF3}. The
$s(M)$--$\delta_c(z)$ relations in the CDM models are steeper than
that given by this simple model, as shown in Figure
\ref{fig:mahssigLCDM123}, because the effective power spectrum
index, which comes into the definition of $s(M)$ increases with the
mass scale.

\begin{figure}
\epsscale{1.0} \plotone{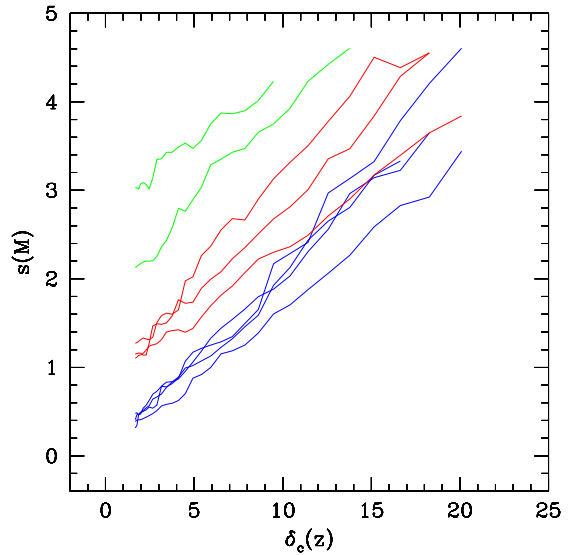} \caption{Simulated mass accretion
histories for individual halos in simulations LCDM1--3. For clarity,
histories of different final masses are presented with different
colors. \label{fig:mahssigLCDM123individual}}
\end{figure}

  Before ending this section, we would like to point out that the
scaling relations obtained here may also apply for individual halos.
As shown in Figure \ref{fig:mahssigLCDM123individual}, the linear
relation between $s(M)$ and $\delta_c(z)$ is also a good
approximation for individual halos, although the slope may change
from halo to halo. Thus, one may model the individual MAHs using a
straight line with its slope changing from halo to halo. The scatter
in the MAHs for halos of a given mass may then be modeled through
the distribution of the slope. Since the linear relation is a good
approximation for different halo masses and in different
cosmologies, as shown in Figures \ref{fig:mahssigSF3} and
\ref{fig:mahssigLCDM123}, this kind of modeling is expected to be
valid for different halos and in different cosmological models. We
will come back to this point in a forthcoming paper.

\section{A universal model for the dark matter halo concentrations}
\label{sec_concentrations}

\subsection{The Mass and Redshift Dependence of Halo Concentration}
\label{sec_cmzrelation}

\begin{figure}
\epsscale{1.0} \plotone{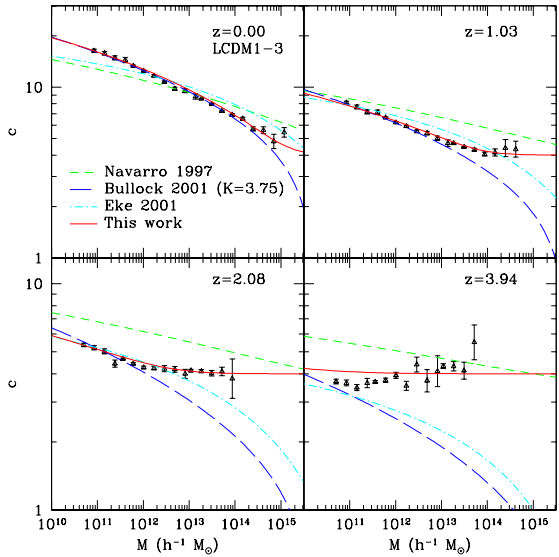} \caption{Predicted and simulated
median concentration as a function of halo mass for simulations
LCDM1--3. Black symbols are results measured in simulations and
error bars the standard deviation among halos of the same mass
divided by the square root of the halo number in the bin, and these
data are in excellent agreement with those presented in Figure 2 of
Zhao et al. (2003b). Solid red lines are our model predictions,
while short-dashed green, long-dashed dark blue, and dot-dashed blue
lines are predictions of the NFW model, the B01 model, and the E01
model, respectively. The B01 model has several versions of model
parameters and here we adopt the version with $F=0.01$ and $K=3.75$.
\label{fig:cmLCDM123}}
\end{figure}

\begin{figure}
\epsscale{1.0} \plotone{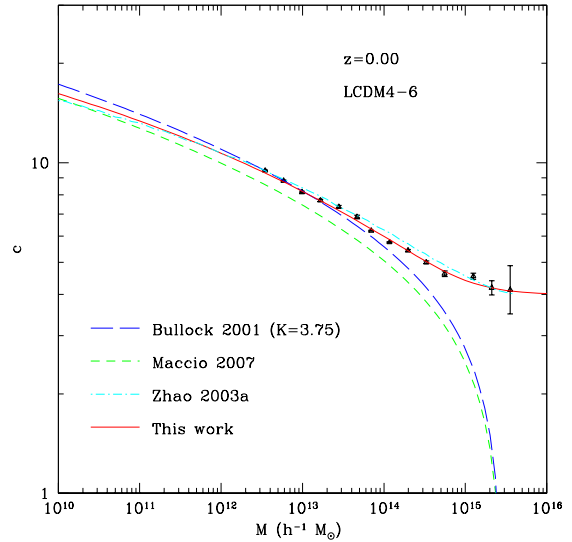} \caption{Predicted and simulated
median concentration as a function of halo mass for simulations
LCDM4--6. Symbols and error bars are the same as in Figure
\ref{fig:cmLCDM123}. Solid red lines are our model predictions,
while short-dashed green, long-dashed dark blue, and dot-dashed blue
lines are predictions of the model of Macci\'o et al. (2007), the
B01 model, and the Z03a model, respectively. \label{fig:cmLCDM456}}
\end{figure}

\begin{figure}
\epsscale{1.0} \plotone{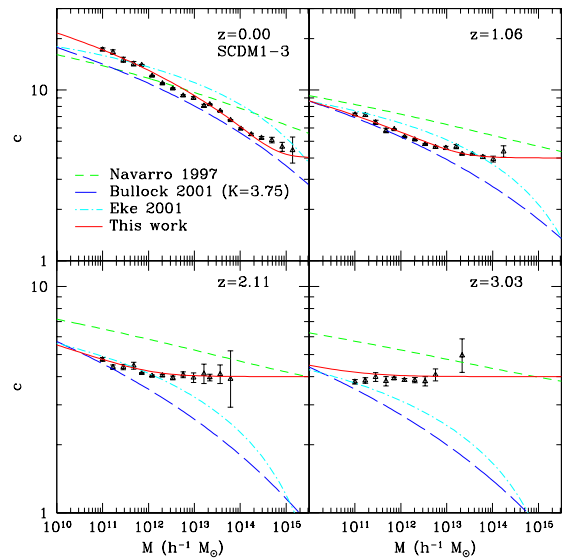} \caption{Predicted and simulated
median concentration as a function of halo mass for simulations
SCDM1--3. Symbols and lines are the same as in Figure
\ref{fig:cmLCDM123}. \label{fig:cmSCDM}}
\end{figure}

 Median values of the halo concentrations measured for the CDM models are
plotted as symbols in Figures \ref{fig:cmLCDM123}--\ref{fig:cmSCDM}.
The results obtained here from simulations LCDM1--3 are in excellent
agreement with the results obtained in Z03b. In these plots, we also
compare our simulation results with the three much-used empirical
models. The first one is the NFW model, which relates the halo
characteristic density at $R_s$ to the universe density at the time
when $50\%$ of the halo mass is already in progenitors of $1\%$ of
the halo mass or bigger. In agreement with B01, our results show
that the NFW model not only fails to predict correctly the redshift
dependence of halo concentration, but also fails to predict the
concentration at $z=0$ accurately. The model of E01 matches our
results better, especially at $z=1$ and $z=2$, but it also fails to
match the concentration--mass relation, particularly at high
redshift. The model of B01 has several versions of model parameters
(see Web site
http://www.physics.uci.edu/{\textasciitilde}bullock/CVIR/). We adopt
the original version with $F=0.01$ and $K=3.75$, where $F$ and $K$
are parameters in Equations (9) and (12) of their
paper.\footnote{Although in the published paper of B01 $K=4$ is
adopted, their latest Web site has deleted all description about
this value and claim that $K=3.75$ should be the original version
corresponding to the published paper.} This version matches well our
simulation results for LCDM1--3 at $z=0$. However, it fails to match
the concentration--mass relation for massive halos, especially at
high $z$. If $F=0.01$ and $K=3.4$ is used, as suggested for total
halo population lately by Macci\'o et al. (2007) and James Bullock
on his Web site, the model underpredicts the concentration by an
amount of $10\%$ even for low-mass halos at $z=0$. The conflicts
between early model predictions and simulation data have already
been discussed by Z03b, and were also confirmed recently by Gao et
al. (2008) who used the Millennium Simulation to carry out an
analysis very similar to that in Z03b. As mentioned in the
introduction, Gao et al. tried some revisions of the NFW model and
found that the revised version still fails to match their simulation
results. The simulation results presented here are in good agreement
with the results in Z03b and in Gao et al. (2008). Since we combine
a set of high-resolution simulations, the dynamical ranges explored
in Z03b and here are larger than the Millennium Simulation. With our
new simulation data of the SCDM model, we find that the B01 cannot
even match our data at $z=0$ unless parameters in their model are
adjusted. As in the LCDM case, the B01 model also fails to account
for the redshift dependence of halo concentration in the SCDM model.

  Z03a found that the scale radius $R_s$ is tightly
correlated with the mass $M_s$ enclosed by it. With this tight
correlation, one can predict halo concentration from the MAHs. Using
the MAH model given in last section, we can predict the halo
concentration according to the Z03a model. The prediction of this
model is shown in Figure \ref{fig:cmLCDM456} as the dot-dashed line.
For comparison, we also show predictions of the B01 model and the
model of Macci\'o et al. (2007) in the same figure. As one can see,
the prediction of the Z03a model is more accurate than the other two
models, particularly for high mass halos.

  Although the model of Z03a matches well the simulation
data, it is not easy to implement. Here we use our simulation
results to find a new model that is more accurate and easier to
implement. Let us first consider an SF case with a given linear
power index. In this case, the halos of a given mass have a uniquely
determined time $t_F$ when their main progenitors reach a fixed
fraction $F$, e.g., $50\%$, of their final mass. There is thus a
one-to-one correspondence between the MAH and $t_F$. If the
concentration of a halo is fully determined by its MAH, one may
build a model to link $t_F$ and the final halo concentration. In
principle, such a relation can be found for $t_F$ at an arbitrarily
fixed fraction of the final halo mass, but the resulting relation
may be very different for scale-free models of different power
indices. What we are seeking is a value of $F$ with which the
relations between $c$ and $t_F$ are the same for all the scale-free
models.

\begin{figure}
\epsscale{1.0} \plotone{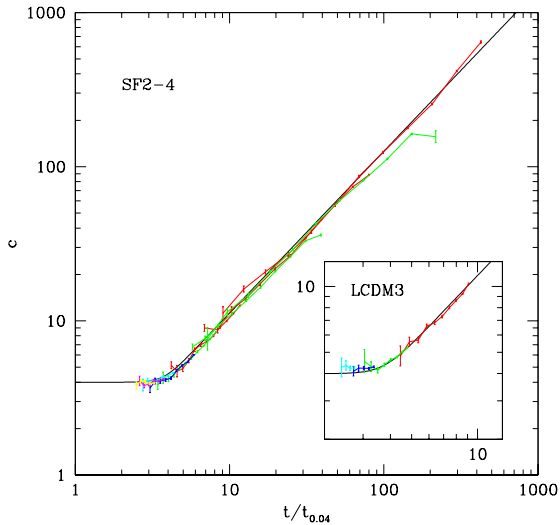} \caption{Concentration of the final
offspring halo as a function of the universe age when its main
progenitor first gained $4\%$ of the final halo mass. Results are
shown for scale-free cases with $n=0,\,-1,\,-2$ and for LCDM3.
Different colors represent different snapshots, and the red line is
for $z=0$. The solid curves are given by Equation (\ref{eq:m2c}).
\label{fig:ct1}}
\end{figure}

After many trials, we find that the concentration of a halo is
tightly correlated with the time $t_{0.04}$ when its main progenitor
gained $4\%$ of its final mass. This relation is almost identical
for $n=0,\,-1$ and $-2$, as shown in Figure \ref{fig:ct1}, but is
slightly lower for the $n=1$ model. Since for realistic cosmological
models, the effective slope of the linear power spectrum on scales
of halos that can form is always less than 0, we exclude the result
of the $n=1$ model in modeling the $c$--$t_{0.04}$ relation. We find
that the relation given by the other three models can be accurately
described by the following simple expression:
\begin{equation}
\label{eq:m2c}
c=\left[4^8+(t/t_{\rm
0.04})^{8.4}\right]^{1/8}=4\times \left[1+(t/3.75 t_{\rm
0.04})^{8.4}\right]^{1/8}\,.
\end{equation}
Nontrivially, this same relation also applies very well to the CDM
models, as shown in the small window of Figure \ref{fig:ct1}.

\begin{figure}
\epsscale{1.0} \plotone{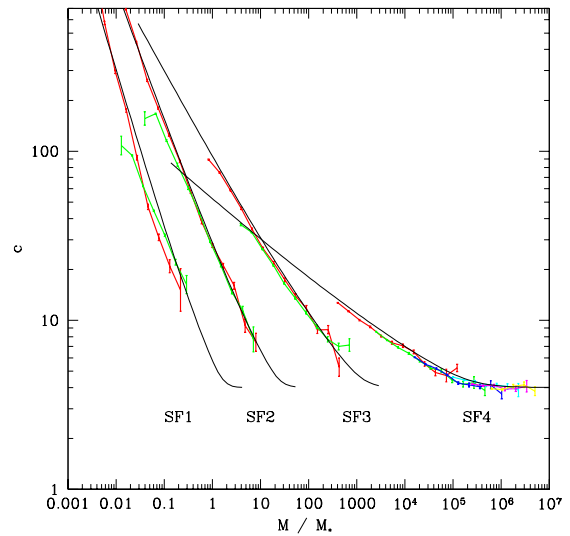} \caption{Halo median concentration
as a function of halo mass for scale-free models with
$n=1,\,0,\,-1,$ and $-2$. Colored symbols are results measured in
simulations and error bars are the standard deviation among halos of
the same mass divided by the square root of the halo number in the
bin. Smooth lines are our model predictions. For clarity, both the
simulation data and the model predictions have been shifted in
horizontal direction by -1, 0, 1.5, and 3.5 for SF1, SF2, SF3, and
SF4, respectively. \label{fig:cmSF}}
\end{figure}

  From our MAH model described in Section \ref{sec_MAHs},
we can easily compute the time $t_{0.04}$ for a halo of mass $M$. We
can then predict its concentration straightforwardly by using
Equation (\ref{eq:m2c}). The concentration--halo mass relations so
obtained are shown as the solid lines in Figure \ref{fig:cmSF} for
the SF models, and in Figures \ref{fig:cmLCDM123}--\ref{fig:cmSCDM}
for the CDM models. Comparing the model predictions with the
corresponding simulation data, we see that our model works
accurately for all the models and for halos at different redshifts.
The typical error of our prediction is less than 5\%. Somewhat
larger deviations ($\sim 10\%$) seen for halos of mass $\sim
10^{11}\msun$ at redshift $z\sim 3$ could be a result of numerical
inaccuracy in the simulations, because these halos are not well
relaxed. Compared with early models, ours is clearly a very
significant improvement. In Appendix B, we describe step-by-step how
to use Equation \ref{eq:m2c} to predict the concentration--mass
relation at any given redshift for a given cosmological model.

 It should be pointed out that both our model and the model
of NFW are based on the correlation between halo concentration and a
characteristic formation time. However, the two models have several
important differences. First, in the NFW model the formation time
was defined as the epoch when half of the halo mass $M$ has been
assembled in its progenitors of masses exceeding $0.01 M$, while in
our model the time is defined as the epoch when its {\it main}
progenitor has gained $4\%$ of the halo mass. Second, the ways to
relate the concentration to the characteristic time are also
different. NFW assumed that the inner density at $R_s$ of a halo is
proportional to the mean density of the universe at the formation
time, while we relate the halo concentration and the characteristic
time by Equation (\ref{eq:m2c}). Finally, NFW used the extended
Press--Schechter formula to compute the formation time, while we use
our model of MAHs to calculate the time $t_{0.04}$. These
differences make a very big difference in the model predictions, as
shown above.

\subsection{The Evolution of Halo Structure Along the Main Branch}
\label{sec_cevolution}

\begin{figure}
\epsscale{1.0} \plotone{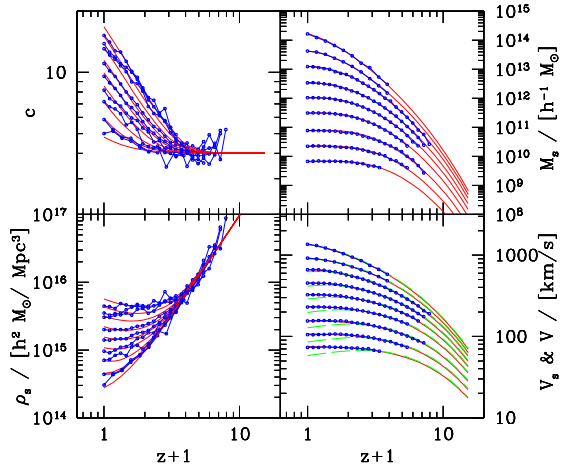} \caption{Predicted (smooth lines)
and simulated (symbols connected with lines) evolution of the median
$c$, $\rho_s$, $M_s$, and $V_s$ for halos of nine different final
masses in simulations LCDM1--3. The increasing direction of final
halo mass among different lines in left/right panels is
downward/upward. Here simulated $\rho_s$, $M_s$, and $V_s$ are
estimated with the simulated median $c$ and the predicted halo mass.
For comparison, evolution of circular velocity at the virial radius,
$V$, is also plotted in the lower right panel as dashed lines.
\label{fig:mshLCDM}}
\end{figure}

  The model described above can also be used to predict
how halo structural properties, such as $c$, $R_s$, $\rho_s$, $M_s$,
and $V_s$, and hence density profile evolve along the main branch.
For a given MAH, $M(z)$, we can estimate $t_{0.04}$ for the current
halo, $M(z)$ [corresponding to the virial radius $R(z)$ and the
circular velocity $V(z)$], at the redshift $z$ in question, and use
the model described above to obtain $c$ at $z$. Since the virial
radius $R$ is determined by $M(z)$, we can then obtain $R_s$,
$\rho_s$, $M_s$, and $V_s$ through $R_s=R/c$, Equations
(\ref{eq:rhosc}) and (\ref{eq:msc}), respectively. The solid lines
in Figure \ref{fig:mshLCDM} show the model predictions in comparison
with the simulation results of LCDM1--3 shown by circles connected
by lines. Clearly, our model also works very well in describing
these evolutions. These results demonstrate that, for a given halo,
our model cannot only predict how its total mass grows with time,
but can also predict how its inner structure changes with time.
Thus, one can plot its NFW density profile at any point of the main
branch, and can obtain the density evolution in spherical shell of
any radius.

 As one can see, for low-mass halos, $\rho_s$, $M_s$, $V_s$,
hence also $R_s$, all remain more or less constant at low redshift,
suggesting that the inner structures of these halos change only
little in the late stages of their evolution. Since by definition
the virial radius $R$ increases with time, the concentration of such
halos increases rapidly with decreasing redshift. Note that the
circular velocity at the virial radius actually decreases with time
for such halos at late time, as shown by the dashed lines in the
lower-right panel of Figure \ref{fig:mshLCDM}. On the other hand,
for massive halos, $\rho_s$, $M_s$, $V_s$, and $R_s$ all change
rapidly with redshift even at $z\sim 0$, implying that the inner
structures of these halos are still being adjusted by the mass
accretion. Therefore, the concentration of such halos increases very
slowly or even remains constant with redshift, much different from
the W02 model which argues that $c \propto a$. All these behaviors
had been illustrated clearly in Z03a and Z03b. As discussed in Z03a,
these different behaviors are mainly due to the fact that massive
halos are still in their early growth phases, which is characterized
by rapid halo growth dominated by major mergers, effectively
reconfiguring and deepening the gravitational potential wells and
causing the collisionless dark matter particles to undergo dynamical
relaxation and to mix up sufficiently to form the core structure,
while low-mass halos have reached the late growth phase, which is
characterized by slower quiescent growth predominantly through
accretion of material onto the halo outskirt, little affecting the
inner structure and potential.

\begin{figure}
\epsscale{1.0} \plotone{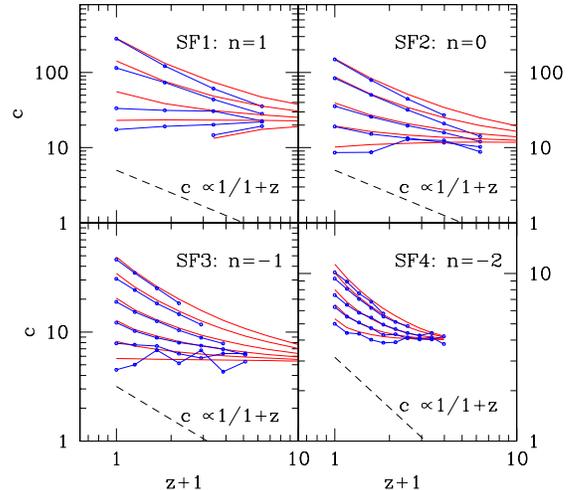} \caption{Predicted (smooth lines)
and simulated (symbols connected with lines) evolution of the median
$c$ for halos of different final masses in simulations SF1--4. The
increasing direction of final halo mass among different lines is
from top to bottom. For comparison, assumption of W02, $c \propto
 1/(1+z)$, is also plotted in each panel as the dashed line. \label{fig:chSF}}
\end{figure}

Figure \ref{fig:chSF} presents the model predictions of the
evolution of median $c$ in comparison with the simulation results of
SF1--4. It is very interesting that the simulated median
concentrations of massive halos in each SF model also remain
constant but the values are not $4$ except for SF4 ($n=-2$). This is
not surprising: as shown in Figure \ref{fig:mahsSF}, the mass
accretion of SF4 halos, which is the fastest in these SF models, is
very similar to the CDM cases and thus triggers dynamical relaxation
and core-structure formation, resulting a constant concentration of
about 4; while for the rest SF models, mass accretion is more or
less slower and even the massive straight-line-MAH halos is in their
slow growth regime ($t/t_{0.04}>3.75$; see Figure \ref{fig:ct1})
and, according to Equation (\ref{eq:m2c}), progenitors on these
straight line MAHs, which have time-invariant $t/t_{0.04}$, will all
have constant but $n$-dependant concentrations. According to Section
\ref{sec_asymptotic}, all median MAHs in SF cases asymptote $M
\propto (1+z)^{-6/(3+n)}$ and, in an Einstein de Sitter universe,
universe age $t \propto (1+z)^{-1.5}$. Combining these two relations
with Equation (\ref{eq:m2c}), we find halo median concentrations
also have a unique asymptotic behavior at high redshift: $c \cong
29.37,\,12.61,\,5.477$, and $4.007$ for $n=1,\,0,\,-1$, and $-2$,
respectively. These numbers are well verified in Figure
\ref{fig:chSF} (except that in the SF1 case $c$ should be shifted a
little bit when compared with the simulation results because a
short-wavelength cutoff in the linear power spectrum have been
included in this simulation) and, for comparison, assumption of W02,
$c \propto 1/(1+z)$, is also plotted in each panel of the figure.
Actually, for this kind of self-similar straight-line MAHs, any
concentration model which claim that halo concentration is tightly
correlated to its MAH should predict a constant concentration, as
required by logic. For the most massive and so very rare halos which
grow faster and faster, as demonstrated in the two upper panels of
Figure \ref{fig:mahsSF}, our model predict a decrease of
concentration with time, as its $t/t_{0.04}$ is diminishing. This
very interesting behavior is again supported by the simulation
results and worth further detailed study.

\subsection{Comparison with the Zhao et al. (2003a) Model}
\label{sec_cevolution}

\begin{figure}
\epsscale{1.0} \plotone{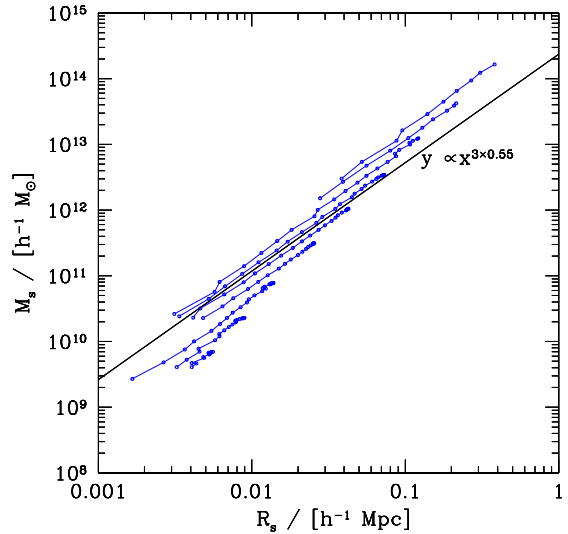} \caption{Inner mass $M_s$ vs. inner
radius $R_s$ along the simulated median MAH obtained from LCDM1--3.
Results are shown for halos with nine different final masses, the
same as those shown in Figure \ref{fig:mshLCDM}.
\label{fig:msrsLCDM}}
\end{figure}

 As mentioned above, the Z03a model for the halo concentrations
is based on the tight correlation between $R_s$ and $M_s$. In Figure
\ref{fig:msrsLCDM} we show the $M_s$--$R_s$ relation obtained from
the simulations used in this paper. Clearly, for a given mass, $M_s$
and $R_s$ are tightly correlated, and the relation is well described
by  $M_s \propto R_s^{3\alpha}$ with $\alpha \simeq 0.55$. The value
of $\alpha$ is between the values $0.48$ and $0.64$ obtained in Z03a
for the late slow and early fast growth regimes, respectively. This
suggests that the model we are proposing here is closely related to
that of Z03a. To show this more clearly, let us start with Equation
(10) in Z03a:
\begin{equation}
\label{eq:cmh}
   {[\ln(1+c)-c/(1+c)]c^{-3\alpha} \over
   [\ln(1+c_0)-c_0/(1+c_0)]c_0^{-3\alpha}}=\left[{\rho(z)\over
   \rho_{0}}\right]^\alpha\left[{M(z)\over
   M_{0}}\right]^{1-\alpha}\,.
\end{equation}
The function of $c$ in square brackets on the left side behaves like
a power law for moderate concentrations and then we will use a power
law function instead. Suppose that we have a halo whose main branch
reaches $4\%$ of its current mass at a time $t_{0.04}$ and assume
that the concentration of the progenitor at $t=t_{0.04}$ is 4.
Substituting quantities of subscript 0 in the above equation with
quantities at $t=t_{0.04}$ gives
\begin{equation}
\label{eq:cmh2}
   {\left(\frac{c}{4} \right)^{\beta-3\alpha}}=\left({\rho\over
   \rho_{0.04}}\right)^\alpha 25^{(1-\alpha)}\,,
\end{equation}
where $\beta \simeq 0.603$ for moderate $c\leq 25$.
We can then obtain
\begin{equation}
\label{eq:cmh3}
   c=4 \left[\left({\rho\over
   \rho_{0.04}}\right)^\alpha 25^{1-\alpha}\right]^{1/(\beta-3\alpha)}
   =4\left[\frac{1}{4}\left({\rho\over
   \rho_{0.04}}\right)^{-1.05/2}\right]\,
\end{equation}
assuming $\alpha=0.55$. This is very similar to relation
(\ref{eq:m2c}) at $t> 3.75 t_{0.04}$, indicating that the model
proposed here is consistent with that of Z03a. Indeed, for a given
MAH, the redshift when $t=3.75\times t_{0.04}$ is uniquely
determined. We found that this redshift separates well the fast
growth regime ($t<3.75 t_{0.04}$) from the slow growth regime
($t>3.75 t_{0.04}$). In the fast growth regime all halos have about
the same median concentration $c\sim 4$ independent of halo mass,
while in the slow growth regime the concentration scales with time
as $\sim 4 (t/3.75 t_{0.04})^{1.05}$. This is exactly the proposal
made in Z03a, Z03b that concentration evolution of a halo depends
much on its mass accretion rate and the faster the mass grows, the
slower the concentration increases.

\section{Conclusions and Discussion}
\label{sec_conclusions}

 A large amount of observations have constrained cosmological
parameters and the initial density fluctuation spectrum to a very
high accuracy. However, cosmological parameters change with time and
the power index of the power spectrum varies with mass scale
dramatically in the so-called concordance $\Lambda$CDM cosmology.
Thus, any successful model for its structural evolution should work
well simultaneously for various cosmological models and different
power spectra.

 In this paper, we use a large set of $N$-body simulations of
various structure formation models to study the MAHs, the mass and
redshift dependence of concentrations and the concentration
evolution histories of dark matter halos. We find that our
simulation results cannot be described by the much-used empirical
models in the literature. Using our simulation results, we developed
new empirical models for both the MAHs and the concentration
evolution histories of dark matter halos, and the latter can also be
used to predict the mass and redshift dependence of halo
concentrations. These models are universal, in the sense that the
same set of model parameters works well for different cosmological
models and for halos of different masses at different redshifts. Our
model predictions also match the PINOCCHIO MAHs very well, which are
automatically output by the Lagrangian semianalytic code PINOCCHIO
without any postprocessing and so are much independent of our
numerical simulations and of our definitions of halo merger trees.
These models are also simple and easy to implement, making them very
useful in modeling the growth and structure of the dark matter halo
population.

 We found that, in describing the median MAH of
dark matter halos, some degree of universality can be achieved by
using $\delta_c(z)$, the critical overdensity for spherical
collapse, to represent the time, and $\sigma (M)$, the rms of the
linear density field on mass scale $M$, to represent the mass. This
is consistent with the Press--Schechter (PS) theory, in which the
dependence on cosmology and power spectrum is all through
$\delta_c(z)$ and $\sigma (M)$.  A more universal relation can be
found by taking into account the local slope of the power spectrum
on the mass scale in consideration. This is again expected, because
the growth of an average halo in a given cosmological background is
determined by the perturbation power spectrum on larger scales.
Indeed, in the extended PS theory, the average mass accretion rate
of a halo of a given mass depends both on $\sigma (M)$ and its
derivative. However, the exact dependence on the shape of the power
spectrum in our empirical model is actually different from that
given by the extended PS theory, as demonstrated clearly by the
discrepancy between our results and the model of van den Bosch
(2002), which is based on the extended PS theory (see Figures
\ref{fig:mahsSF}--\ref{fig:mahsSCDM12OCDM}). We find that another
modification is required in order to correct the deviation of halos
corresponding to relatively low peaks. We suggest that such
deviation is due to the neglect of the large-scale tidal field on
the formation of dark matter halos, as found in Wang et al. (2007).
Thus, the empirical model we find for the halo accretion histories
may help us to develop new models for the mass function and merger
trees of dark matter halos. In addition, we find that a linear
relation between $s(M)$ [defined in Equation (\ref{eq:sasM})] and
$\delta_c(z)$ is also a good approximation for individual halos,
although the slope may change from halo to halo. Thus, the MAHs of
individual halos may be modeled with a set of straight lines, in the
$s(M)$--$\delta_c(z)$ space, with different slopes specified by a
distribution function. Research along these lines is clearly
worthwhile, and we plan to come back to these problems in
forthcoming papers.

Our other finding is that the concentration of a halo is strongly
correlated with $t_{0.04}$, the age of the universe when its
progenitor on the main branch first reaches $4\%$ of its current
mass. This is consistent with the general idea that the structure of
a dark matter halo is correlated with its formation history. The
concentration of a halo selected at a given redshift is determined
by the ratio of the universe age at this redshift, $t$, to
$t_{0.04}$. If this ratio is smaller than 3.75, the halo is in the
fast growth regime and its concentration is expected to be $c\sim
4$, independent of its mass. For a halo in the fast growth regime,
its first $4\%$ mass settled down at a late epoch (of age $>t/3.75$)
and so has low density. In addition, its current accretion rate is
large in the sense that it only takes a time interval of much less
than $2.75t/3.75$ to gain the 96 percent of mass remaining. Such
accretion is expected to have a significant impact on the inner
structure that is not very dense, and so the inner structure of this
halo is adjusted constantly by the mass accretion. In contrast, for
a halo in the slow growth regime ($t/t_{0.04}>3.75$), its first
$4\%$ mass settled down at an early epoch when the universe age was
smaller than $t/3.75$, and so this fraction of the mass is expected
to settle into a very dense clump. Also, for such a halo, the
current accretion rate is small in the sense that it takes a time
interval of more than $2.75t/3.75$ to gain the $96\%$ of mass
remaining. Such slow accretion is not expected to have a significant
impact on the inner dense structure that has already formed. All
these are in qualitative good agreement with the results presented
in Z03a. Our model can also be used to predict accurately how halo
structural properties, such as $c$, $R_s$, $\rho_s$, $M_s$ and
$V_s$, and hence its density profile evolve along the main branch,
making it very useful for our understanding of the formation and
evolution of dark matter halos. Assuming that the NFW profile is a
good approximation for the halo density profile, the model can
predict too the MAHs, the mass and redshift dependence of
concentrations and the concentration evolution histories for
different halo definitions. In a forthcoming paper, we will apply
this model to individual halos.

We found that in a self-similar cosmology halos on some
characteristic mass scale have a straight-line median MAH and a
time-invariant median concentration. All halos asymptote these
median features at high redshift: halos under this mass scale will
grow slower and slower on average and their concentration will
increase with time; while halos above this mass scale will grow
faster and faster on average and their concentration will decrease.
On the other hand, in $\Lambda$CDM and OCDM cases, both the decrease
of cosmological density parameter $\Omega_m$ with time and the
increase of power index of the power spectrum with mass scale will
slow down the halo mass growth rate, inducing that at the present
time halos on nearly all mass scales show two distinct phases in
their mass growth histories, as found by Z03b and Z03a. Z03a and
Z03b also pointed out that when halo mass accretion rate reaches
some criteria, the fast mass growth will trigger dynamical
relaxation and core-structure formation, resulting in a constant
concentration of about 4. Here we found another mechanism that will
lead to a time-invariant concentration along the main branch of a
halo merger tree: a self-similar straight-line MAH will induce a
constant concentration, as required by logic, and when the mass
accretion rate is below the above criteria, the value of this
constant is determined by the accretion rate and is always larger
than 4.

\acknowledgments

We are grateful to Simon D. M. White for helpful discussion and to
Julio Navarro and James Bullock for providing their codes to compute
$c$. The work is supported by NSFC (10303004, 10533030, 10873028,
10821302, 10878001), by 973 Program (2007CB815402, 2007CB815403),
Shanghai Qimingxing project (04QMX1460), and by the Knowledge
Innovation Program of CAS (KJCX2-YW-T05).  H.J.M. thanks the Chinese
Academy of Sciences for supporting his visit to SHAO. He also
acknowledges the support of NSF AST-0607535, NASA AISR-126270 and
NSF IIS-0611948.  Part of the numerical simulations presented in
this paper were carried out at the Astronomical Data Analysis Center
(ADAC) of the National Astronomical Observatory, Japan.

\appendix

In the appendices A and B, we give a detailed user's guide on how to
use the results of the present paper to compute the median mass
accretion history and the concentration of dark matter halos for a
given cosmological model.

\section{Calculating the median mass accretion history of halos}

For a given cosmological model, one can calculate
$\delta_c(z)=\delta'_c [\Omega_m(z),\Omega_\Lambda(z)]/D(z)$ for a
given redshift $z$, where $\delta'_c[\Omega_m(z),\Omega_\Lambda(z)]$
is the conventional spherical collapse threshold at redshift $z$,
$D(z)$ is the linear growth factor normalized to 1 at redshift
$z=0$, and $\delta_c(z)$ is the collapse threshold for the density
field linearly evolved to $z=0$. Given a linear power spectrum at
$z=0$, one can also calculate the rms density fluctuation
$\sigma(m)$ for a spherical volume of mass $m$, as well as $s(m)$.
Once these quantities are obtained, one can calculate the median
mass growth history of dark matter halos of mass $M_{\rm obs}$ at
redshift $z_{\rm obs}$, i.e., the median mass $M(z|z_{\rm obs},
M_{\rm obs})$ of their main progenitors at a higher redshift
$z>z_{\rm obs}$. The step-by-step procedure is as follows.

\begin{enumerate}
\item
With $\sigma(M_{\rm obs})$, $s(M_{\rm obs})$ and $\delta_c(z_{\rm
obs})$ for halos of mass $M_{\rm obs}$ at redshift $z_{\rm obs}$,
one can calculate $w(z_{\rm obs}, M_{\rm obs})$ according to
Equation (\ref{eq:wzm}) and the shift $p(z_{\rm obs}, z_{\rm obs},
M_{\rm obs})$ according to Equation (\ref{eq:pzobs}). Let $z=z_{\rm
obs}$ and $M=M_{\rm obs}$ temporarily.
\item
One can obtain the accretion rate $d\lg\sigma(M)/{d\lg\delta_c(z)}$
at this redshift by substituting the above $w$ and $p$ into Equation
(\ref{eq:accretionrate}).
\item
At an incrementally higher redshift $z'=z+\Delta z$, one can
calculate the new critical collapse threshold $\delta_c(z')$ at
$z'$, and its change $\Delta\lg\delta_c(z)$ between $z'$ and $z$,
and hence obtain the change in $\lg\sigma (M)$ through $\Delta
\lg\sigma (M)=d\lg\sigma(M)/{d\lg\delta_c(z)} \times
\Delta\lg\delta_c(z)$. One can then get a new $\sigma (M')$, with
$M'$ being the median mass of the main progenitors at $z'$.  The
mass $M'$ can be obtained from $\sigma$ through the function $\sigma
(m)$.
\item
In order to trace the history further backward, one needs to
calculate $w(z',M')$ according to Equation (\ref{eq:wzm}) and
calculate the shift $p(z', z_{\rm obs}, M_{\rm obs})$ according to
Equation (\ref{eq:pz}). Let $z=z'$, $M=M'$, and repeat steps 2 and
3.
\end{enumerate}
Then step by step, one can trace the MAH backwards to high redshift.
One can also calculate the halo radius $R(z)$ and halo circular
velocity $V(z)$ along this MAH according to Equations (\ref{eq:mh}
and \ref{eq:vh}) by setting $\Delta_h=\Delta_{\rm vir}$ as defined
in Section 1.

\section{Calculating the halo concentration}

For a given MAH ending at $[z_{\rm obs}, M_{\rm obs}]$ in a given
cosmology, one can predict concentrations for {\it all the main
progenitors at different redshifts} in addition to the final
offspring, i.e., one can obtain the concentrations along the MAH,
$c(z|z_{\rm obs}, M_{\rm obs})$. Then many other useful quantities
can be calculated as shown below. There are two steps in computing
the concentrations.

\begin{enumerate}
\item
From a given MAH, $M(z|z_{\rm obs}, M_{\rm obs})$, one can compute
$z_{0.04}$, the redshift at which the main progenitor of the main
progenitor halo at $z(\geq z_{\rm obs})$ grows to $0.04M$, where $M$
is the mass of the main progenitor halo at $z$.
\item
In the given cosmology, one can calculate the ages of the universe,
$t$ and $t_{0.04}$, at these two redshifts, $z$ and $z_{0.04}$, and
obtain $c(z|z_{\rm obs}, M_{\rm obs})$ by substituting them into
Equation (\ref{eq:m2c}).
\item
Furthermore, one can calculate the characteristic inner quantities
of the halo along the MAH, such as $R_s(z|z_{\rm obs}, M_{\rm
obs})$, $\rho_s(z|z_{\rm obs}, M_{\rm obs})$, $M_s(z|z_{\rm obs},
M_{\rm obs})$ and $V_s(z|z_{\rm obs}, M_{\rm obs})$, according to
the definition of $c$ and Equations (\ref{eq:mh}--\ref{eq:msc}), and
hence can plot the NFW density profile at any point of the MAH
according to Equation (\ref{eq:nfw}).
\end{enumerate}

In the literature, there are several definitions for halo radius.
Assuming that the NFW profile is a good approximation for the halo
density profile, one can predict the MAHs and the concentration
evolution histories for different halo definitions, because halos of
different definitions have the same inner structures (such as $R_s$,
$\rho_s$ and $M_s$) although their boundaries are different. For
example, according to Equations (\ref{eq:rhosc} and \ref{eq:msc}),
one can calculate $c_{\rm 200c}(z|z_{\rm obs},M_{\rm obs})$, $M_{\rm
200c}(z|z_{\rm obs}, M_{\rm obs})$ , $V_{\rm 200c}(z|z_{\rm obs},
M_{\rm obs})$ for halo definition $\Delta_h(z)=200/\Omega_m(z)$ and
one can also calculate $c_{\rm 200m}(z|z_{\rm obs}, M_{\rm obs})$,
$M_{\rm 200m}(z|z_{\rm obs}, M_{\rm obs})$, $V_{\rm 200m}(z|z_{\rm
obs}, M_{\rm obs})$ for halo definition $\Delta_h(z)=200$.

Note that $c(z|z_{\rm obs}, M_{\rm obs})$ obtained with steps 1 and
2 is the concentration of a main progenitor of a halo of mass
$M_{\rm obs}$ at $z_{\rm obs}$. As we pointed out in Section 3.2,
this population of main progenitors of mass $M$ at $z$ is not
statistically the same as the whole population of halos of mass $M$
at $z$, although in some cases the difference between them is
negligible. The former population resides in environment of more
frequent mergers and so has slightly smaller concentration. In any
case, one can easily compute $c(M,z)$ for the whole population of
halos of mass $M$ at $z$ if one sets $z_{\rm obs}=z$ and $M_{\rm
obs}=M$ in the above calculation.

\section{A calculator, a user-friendly code and tables available on internet}

 A calculator which allows one to interactively generate the median
MAHs, the concentration evolution histories and the mass and
redshift dependence of concentrations of dark matter halos is
provided at {\tt http://www.shao.ac.cn/dhzhao/mandc.html}, together
with a user-friendly code to make the relevant calculations. The
calculator and the code can be used for almost all current
cosmological models with or without inclusion of the effects of
baryons in the initial power spectrum. We also provide tables for
median concentrations of halos with different masses at different
redshifts in several popular cosmological models. Detailed
instructions on how to use the calculator, the code and the tables
can be found at the Web site.

\end{document}